\documentstyle[12pt]{article}
\input psfig.sty
\def\lsim{\mathrel{\rlap{\lower4pt\hbox{\hskip1pt$\sim$}}
    \raise1pt\hbox{$<$}}}         %less than or approx. symbol
\def\gsim{\mathrel{\rlap{\lower4pt\hbox{\hskip1pt$\sim$}}
    \raise1pt\hbox{$>$}}}         %greater than or approx. symbol
\title{Neutrino Physics}
\author{Wick C. Haxton\\
Institute for Nuclear Theory, Box 351550, \\ and Department of Physics, Box
351560,\\
University of Washington,
Seattle, WA 98195\\
and\\
Barry R. Holstein\\
Institut f\"{u}r Kernphysik\\
Forschungszentrum J\"{u}lich\\
D-52425 J\"{u}lich, Germany\\
and\\
 Department of Physics and Astronomy\\
University of Massachusetts,
Amherst, MA  01003}
\begin{document}
\begin{titlepage}
\maketitle
\begin{abstract}
The basic concepts of neutrino physics are presented at a level
appropriate for integration into elementary courses on quantum mechanics
and/or modern physics.
\end{abstract}
\vfill
\end{titlepage}
\section{Introduction}
The neutrino has been in the news recently, with reports that
the SuperKamiokande collaboration --- which operates a 50,000
ton detector of ultrapure water isolated deep within the
Japanese mine Kamiokande --- has found evidence
of a nonzero neutrino mass \cite{time}.
The neutrino, a ghostly particle which can easily pass through
the entire earth without interacting, has long fascinated
both the professional physicist and the layman, as this ditty
from writer John Updike \cite{updike} attests
\begin{tabbing}
        tabtry  \=        tabtry                            \kill
                \>Neutrinos, they are very small\\
                \>They have no charge and have no mass\\
                \>And do not interact at all.\\
                \>The earth is just a silly ball\\
                \>To them, through which they simply pass,\\
                \>Like dustmaids down a drafty hall\\
                \>Or photons through a sheet of glass.\\
                \>They snub the most exquisite gas,\\
                \>Ignore the most substantial wall,\\
                \>Cold-shoulder steel and sounding brass,\\
                \>Insult the stallion in his stall,\\
                \>And, scorning barriers of class,\\
                \>Infiltrate you and me! Like tall\\
                \>And painless guillotines, they fall\\
                \>Down through our heads into the grass.\\
                \>At night, they enter at Nepal\\
                \>And pierce the lover and his lass\\
                \>From underneath the bed---you call\\
                \>It wonderful; I call it crass.\\
\end{tabbing}
We present this pedagogical discussion of basic neutrino physics
in the hope that aspects of this topical and fascinating
subject can be integrated into introductory courses, providing
a timely link between classroom physics and science news in
the popular press.  In this way an instructor may be able to build
on student curiosity in order to enrich the curriculum with some
unusual new physics.  In this spirit we present below some of
the basic physics underlying massive neutrinos and neutrino
mixing, as well as other properties of neutrinos relevant to
both terrestrial experiments and astrophysics.

\section{Neutrinos: History}

We begin with a bit of history---an interesting and more detailed
discussion can be found in Laurie Brown's article in the September 1978
issue of Physics Today \cite{brown}.  Nuclear beta decay is a
natural form of radioactivity wherein a parent nucleus decays
to a daughter with the same atomic mass, but an atomic number
changed by one unit, with the missing charge carried off by
an electron or positron
\begin{equation}
(A,Z)\rightarrow (A,Z \pm 1) + e^{\mp} \label{eq:aa}
\end{equation}
This is quite literally nuclear transmutation of the type that
fascinated alchemists of an earlier age.  One well-known
example is the decay of a free neutron into a proton and
electron, with a half life of about 10 minutes.  Another
is the decay of a bound neutron in tritium to produce an electron
and ${}^3$He with a half life of 12.26 years:
the effects of the nuclear binding in changing the energy
released in the decay is responsible for the great increase
in the half life.
At the end of the 1920's the existence
of such beta emitters was well established.  However,
the spectrum of the emitted electrons was puzzling.
If beta decay occurs
from rest into a two-body final state as given in Eq. \ref{eq:aa},
momentum conservation would require the momenta of the emitted electron and
recoiling nucleus to be equal and opposite.  Energy conservation would
then fix the outgoing electron energy which, because the nucleus
is heavy and thus recoils with a negligible velocity, is
nearly equal to the difference of the parent and daughter
nuclear masses (known as the reaction energy release or Q-value)
\begin{equation}
Q\simeq M(A,Z)-M(A,Z \pm 1)
\end{equation}
As the Q-value in the beta decay of tritium is 18.6 keV,
one would expect a monochromatic spectrum with all emitted electrons having
this
energy.  Instead experimentalists found a continous spectrum of electron
energies
ranging from the
rest mass $m_e$ to the Q-value, peaking at an energy
about halfway in between, as shown in Figure 1.  Various explanations
were considered---Niels Bohr even proposed the possibility that
energy conservation was no longer exact in such subatomic processes,
and rather preserved only in a statistical sense, somewhat in analogy
with the second law of thermodynamics!
However, in a letter dated December, 1930, Pauli suggested an
alternative explanation---that an unobserved light
neutral particle (called by him the ``neutron'' or neutral
one but later renamed by Fermi the ``neutrino'' or little
neutral one) accompanied the outgoing electron and carried off the
missing energy that was required to satisfy energy conservation.
Pauli offered this explanation tentatively as a ``desperate
remedy'' to solve the energy problem.  Although he publicized it in
various talks over the next three years, no publication occurred until his
contribution to the Seventh Solvay
Conference in October 1933 \cite{pauli}.  He also proposed (correctly)
that the neutrino was a particle carrying spin 1/2 in order to
satisfy angular momentum conservation and statistics.

\begin{figure}
\psfig{bbllx=0cm,bblly=4.0cm,bburx=12cm,bbury=16cm,figure=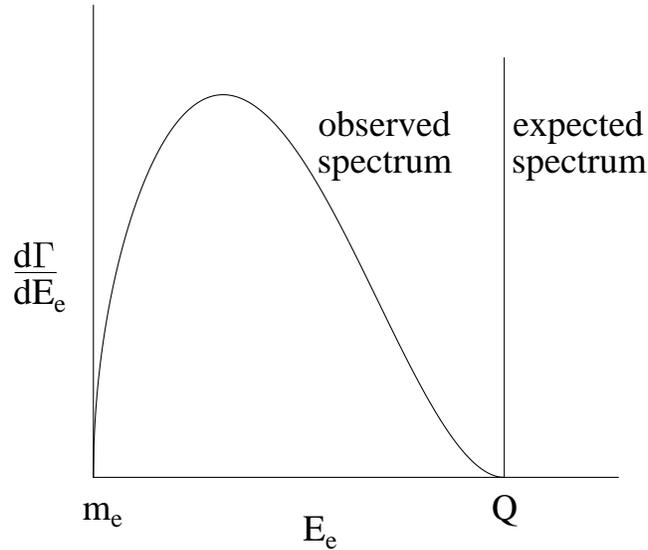,height=3.3in}
\caption{The $\beta$ energy spectrum for decay into a heavy
daughter nucleus, electron, and neutrino is compared to the
monoenergetic spectrum for decay into a daughter nucleus 
and electron, only.  The spectrum is idealized: distortions due
to the Coulomb interaction between the electron and daughter
nucleus have been neglected.}
\end{figure}
  
Fermi was present at a number of Pauli's presentations and discussed the
neutrino with him on these occasions.  In 1934 he published his
insightful model for the beta decay process, and indeed
for weak interactions in general \cite{fermi}.
He described beta decay in analogy with Dirac's
successful model of the electromagnetic interaction, wherein
two charged particles interact via the exchange of a (virtual) photon
that is produced and then absorbed by the electromagnetic currents
associated with the particles ({\it cf.} Figure 2a).
Fermi represented the weak interaction in terms of the product of weak
``currents,'' one connecting the initial and final nucleon and the
other connecting the final state electron/positron and Pauli's
neutrino ({\it cf.} Figure 2b). In electromagnetism the virtual photon connects
the
two currents at distinct points in space-time: indeed the masslessness
of the photon is the reason for the long-range Coulomb force.
In his weak interaction theory, however, Fermi connected the currents
at the {\it same} space-time point, in effect assuming that the weak
interaction is very short ranged.  The strength of the interaction
was determined by an overall coupling strength $G_F$
\begin{equation}
{\cal H}_w={G_F\over \sqrt{2}}\psi^\dagger_pj_\mu\psi_n
\psi^\dagger_ej^\mu\psi_\nu\label{eq:bb}
\end{equation}
As written above, one of the weak currents is associated with
the conversion of a neutron into a proton, and the other with
the production out of the vacuum of an electron and antineutrino,
which carry off almost all of the released energy.
The electron spectrum predicted by this weak Hamiltonian can be
readily calculated by using Fermi's golden rule for the differential
decay rate, yielding in the no-nuclear-recoil approximation
\begin{eqnarray}
d\Gamma& \sim &\left({G_F\over \sqrt{2}}\right)^2
{d^3p_e\over (2\pi)^3}{d^3p_\nu\over (2\pi)^3}
2\pi\delta(Q-E_e-E_\nu)|M_w|^2\nonumber\\
&\sim& \left({G_F\over \sqrt{2}}\right)^2
{(4\pi)^2\over (2\pi)^5}dE_eE_ep_e\sqrt{(Q-E_e)^2-m_\nu^2}(Q-E_e)|M_w|^2
\end{eqnarray}
where $M_w$ is the nuclear matrix element.
Taking $m_\nu=0$ and assuming $|M_w|^2\sim$ constant we find
\begin{equation}
{d\Gamma\over dE_e}\sim p_eE_e(Q-E_e)^2 \label{eq:ttt}
\end{equation}
where $p_e$ and $E_e$ are the momentum and energy of the electron.
The excellent fit to experimentally measured spectra was an
important confirmation of Fermi's theory, and thus of Pauli's
postulate of the neutrino.  Yet it would take another two
decades to detect this elusive particle directly.

\begin{figure}
\psfig{bbllx=0.0cm,bblly=4.1cm,bburx=13cm,bbury=14.6cm,figure=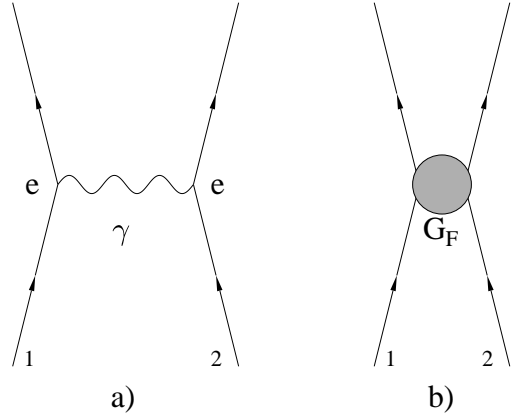,height=2.5in}
\caption{Simple diagrammatic representation of a) the electromagnetic
interaction arising from the exchange of a virtual photon between
the electromagnetic currents generated by two charge particles and
b) the weak interaction as a local product of two weak currents.}
\end{figure}

The reason behind the difficulty of direct detection of the neutrino can
be seen in the size of the weak coupling $G_F$.
From the 887 second lifetime of the neutron one finds that
$\Gamma_n = \hbar/\tau_n \sim 7 \times 10^{-28}$ GeV, while
Fermi's theory gives
\begin{eqnarray}
\Gamma_n&\sim& |{G_F\over
\sqrt{2}}|^2\int{d^3p_e\over
(2\pi)^3}{d^3p_\nu\over
(2\pi)^3}2\pi\delta(m_n-m_p-E_e-E_\nu) |M_w|^2\nonumber\\
&=& {G_F^2\over 2}{(4\pi)^2\over (2\pi)^5}\int_{m_e}^{m_n-m_p}
dE_eE_ep_e(m_n-m_p-E_e)^2 |M_w|^2\nonumber\\
&\simeq&4.59\times 10^{-19}\,{\rm GeV}^5 G_F^2 |M_w|^2
\end{eqnarray}
As $|M_w|^2 \sim 6$, one finds $G_F \sim 10^{-5}$ GeV$^{-2}$.
In order to understand why this interaction is called
``weak'', we note that the contemporary picture is that $G_F$ can
be understood in analogy with the electromagnetic interaction as the result
of an exchange of a virtual but very massive and charged
``photon"---the W-boson---so that $G_F \sim e^2/M_W^2$ with $M_W \sim 80$ GeV.  By
the
uncertainty principle, however, a virtual particle with such a heavy
mass can propagate a distance $\Delta x\sim c \Delta t \sim \hbar c/M_W
\sim 0.002$ fm., {\it i.e.} a very small fraction of the nucleon radius.
It is this exceeding small interaction range --- which allows
point particles, such as the neutrino and the quarks that are
the underlying consituents of a nucleon, to interact only if
they fortuitously pass very close
to one another --- that is responsible for the weakness of
the weak interaction.

Fermi's theory is a relativistic quantum field theory,
wherein a given field operator $\psi_q$ represents both an
annihilation operator for generic particle $q$ as well as a creation
operator for the corresponding antiparticle $\bar{q}$.  Thus Eq.
\ref{eq:bb} (and its generalization for beta decay of protons
within the nucleus) contains not only the interaction for beta
decay---$n\rightarrow pe^-\bar{\nu}$ and $p\rightarrow ne^+\nu$---but also
\begin{itemize}

\item [i)] electron capture, where an atomic electron orbiting the nucleus
interacts with one of the nuclear protons, converting it to a
neutron and producing an outgoing neutrino, $e^- + p \rightarrow
n + \nu$.  (An analogous process for positrons, $e^+ + n \rightarrow
p + \bar{\nu}$, is important in the hot plasmas encountered in
the big bang and in explosive stellar environments.)

\item [ii)] The charged current neutrino reactions $\nu + n \rightarrow
p + e^-$ and $\bar{\nu} + p \rightarrow n + e^+$ (which are the
inverses of the reactions in i), but are often referred to as
``inverse beta decay").

\item [iii)] The exotic resonant reactions $\bar{\nu} + e^- + p
\rightarrow n$ and $\nu + e^+ + n \rightarrow p$, the true
inverse reactions of beta decay.  The first can occur
in an atom; both can take place in astrophysical plasmas.

\end{itemize}

Reaction ii) is the one relevant for neutrino detection, as
the produced electron/positron and nuclear transformation are
signals for a neutrino interaction.  Such a process
is characterized by the scattering cross section $\sigma$, which
when multiplied by the incident particle flux and the number of
scattering targets and integrated over solid angles,
yields the number of scattering events per unit time.  The cross
section has the dimensions of area.  In the
approximation that the neutron is much heavier than the positron
\begin{equation}
\sigma\sim |{G_F\over \sqrt{2}}|^2\int {d^3p_e\over (2\pi)^3}
2\pi\delta(m_p+E_\nu-m_n-E_e)|M_w|^2 \sim {G_F^2\over 2\pi}p_eE_e
|M_w|^2
\end{equation}
For incident neutrino energies large compared to the electron rest
mass this becomes $\sigma\sim G_F^2E_\nu^2 |M_w|^2/2\pi \sim
10^{-44}$ cm${}^2$ for $E_\nu \sim$ 1 MeV.  If we
consider a single neutrino passing through a slab of material having
a target density $\rho$ ({\it e.g.}, $\sim 10^{23}$ atoms/cm${}^3$ for
typical materials), it would travel a distance
\begin{equation}
\Delta x\sim 1/(\rho\sigma) \sim 10^{21} \mathrm{cm}
\end{equation}
before interacting, a distance equivalent to 100 billion earth radii!
This is indeed a {\it weak}
interaction!
The only way to circumvent this problem is to have lots of low energy
neutrinos.  The
orginal plan of a Los Alamos team led by Fred Reines and Clyde Cowan
was to use a fission bomb to produce the needed neutrinos.  Later, however,
they decided to use a nuclear reactor, which produces large
numbers of antineutrinos.
Working at the Savannah River reactor, in South Carolina, which has a neutrino
flux of
about $10^{13}$ per square centimeter per second,
they designed a detector consisting of two plastic tanks each filled
with 200 liters of water, in which was dissolved cadmium chloride.
The protons in the water provided the target for the reaction $\bar{\nu} + p
\rightarrow
n + e^+$, while the cadmium has a large cross section for neutron
capture.  The tanks were sandwiched between three scintillation
detectors.  The group looked for a signal consisting of
gamma rays from the annihilation of the emitted positron
on an electron followed closely (within a few microseconds)
by gammas from the deexcitation of the cadium nucleus that had
captured the neutron.
The observed signal was correlated with the
reactor being in operation.   With this evidence they announced in 1956
that the neutrino had been detected \cite{disc}, almost 25 years after
Pauli's original suggestion.  In 1996 Reines was awarded the Nobel
Prize for this discovery.  (Clyde Cowan had died years earlier.)

However, this was not the end of the story, but only the beginning.
Indeed within seven years there was another Nobel-Prize-winning
neutrino discovery: Lederman, Schwartz, and Steinberger demonstrated
that there was more than one type of neutrino \cite{led}.
In order to explain their discovery we first provide a bit of theoretical
background.  As discussed above, the processes associated with
Fermi's picture of beta decay are called ``weak,''
characterized by rates or cross sections nearly twenty orders
of magnitude smaller than those involving strongly interacting
particles, such as the cross sections for scattering one nucleon
off another.  Particles, such as neutrinos, that do not participate
in strong interactions are called ``leptons.''  Thus the
electron is also a lepton.  (Of course, as
the electron carries a charge, it has both electromagnetic and
weak interactions, while we believe neutrinos react only weakly.)
In the 1930's another charged particle was found which does not interact
strongly---the muon.  Except for the fact that it is about
200 times heavier, the
muon's behavior is remarkably similar to that of the electron \cite{rabi}.
The muon, however, is unstable and decays in $\sim 10^{-6}$ sec
into an electron and two neutral unseen particles
that we now know are a neutrino and an antineutrino.  This timescale
is appropriate for a weak interaction, as can be seen from the
estimate
\begin{eqnarray}
\Gamma_\mu&=&{1\over \tau_\mu}\sim|{G_F\over
\sqrt{2}}|^2\int{d^3p_e\over (2\pi)^3}{d^3p_\nu\over
(2\pi)^3}{d^3p_{\bar{\nu}}\over (2\pi)^3}
(2\pi)^4\delta^4(p_\mu-p_e-p_\nu-p_{\bar{\nu}})\nonumber\\
&\sim&{G_F^2m_\mu^5\over 1636\pi^3}\quad i.e.\quad
\tau_\mu \sim 10^{-5}\,{\rm sec}
\end{eqnarray}
Muon decay fits easily into Fermi's interaction
provided the lepton current is generalized to
\begin{equation}
J_\mu^{\rm lep}=\psi^\dagger_ej_\mu\psi_\nu+\psi_\mu^\dagger j_\mu\psi_\nu
\label{eq:zz}
\end{equation}
Then the product of the lepton current with its hermitian conjugate
yields an interaction responsible for muon decay.

In order to understand the experiment of ref. \cite{led}, it is useful
to go one step further and introduce modern quark notation.  In
the quark model the neutron is a composite object comprised of a pair of d
quarks
and a single u quark, while the proton consists of a pair of u quarks
and a single d quark.  The weak current connecting the proton and neutron can
then
be replaced by a corresponding current connecting a u quark with a d quark,
\begin{equation}
J_\mu^{\rm had}=\psi_n^\dagger j_\mu\psi_p\rightarrow\psi_d^\dagger j_\mu\psi_u
\end{equation}
As the field operator $\psi_d^\dagger$ can both create a $d$ quark and
destroy the corresponding antiparticle ($\bar{d}$), this same current
describes the quark component of the process where a $\bar{d}u$ system
({\it i.e.} a $\pi^+$ meson) decays to a muon and a neutrino
\begin{equation}
\pi^+\rightarrow\mu^++\nu
\end{equation}
which is the dominant decay mode of the charged pion \cite{ftnt}.
The experimenters of ref. \cite{led} collided neutrinos
from such decays with neutrons in an
attempt to produce electrons {\it and} muons,
as predicted by the current of Eq. \ref{eq:zz}.
But they found only muons, {\it not} electrons.  The
explanation for this result is that neutrinos come in {\it two} distinct
species, an
electron type $\nu_e$ and a muon type $\nu_\mu$, with the
weak current coupling electrons only to $\nu_e$ and muons only to $\nu_\mu$
\begin{equation}
J_\mu^{\rm
lep}=\psi_e^{\dagger} j_\mu\psi_{\nu_e}+\psi_\mu^{\dagger} j_\mu\psi_{\nu_\mu}\label{eq:dd}
\end{equation}
The neutrino produced in pion decay thus must be a $\nu_\mu$
and of the wrong type, or ``flavor," to produce an electron.
In 1977 a third charged lepton, the $\tau$, was discovered and
another term has now been added to this equation---the coupling
of the $\tau$ to its neutrino, the $\nu_\tau$.
Measurements of the decay width of the neutral
$Z$-boson \cite{zb} and astrophysical arguments based on the helium
abundance in the universe \cite{sch} suggest that this may
exhaust the set of lepton-neutrino pairs: there appear to be
no more light neutrinos beyond the $\nu_\tau$.

The modern picture of
the weak interaction consists not only of three doublets of charged
lepton-neutrino pairs but also of three doublets (often called
``generations") of charge 2/3, charge -1/3 quarks---(u,d),(c,s),(t,b).
The charged weak current then can be written as the sum of
six separate currents connecting such quark and lepton doublets
\begin{equation}
J_\mu=J_\mu^{\rm had}+J_\mu^{\rm lep}=
\left(\psi^\dagger_d\,\,\psi^\dagger_s\,\,\psi^\dagger_b\right)U_{KM}j_\mu
\left(\begin{array}{l}
\psi_u\\
\psi_c\\
\psi_t
\end{array}\right)+\left(\psi_e^\dagger\,\,\psi_\mu^\dagger\,\,
\psi_\tau^\dagger\right)j_\mu
\left(\begin{array}{l}
\psi_{\nu_e}\\
\psi_{\nu_\mu}\\
\psi_{\nu_\tau}
\end{array}\right)\label{eq:gg}
\end{equation}
Low-energy weak interactions are then described by a effective
current-current interaction with a single overall coupling $G_F$
\begin{equation}
{\cal H}_w={G_F\over \sqrt{2}}J_\mu^\dagger J^\mu
\end{equation}
Such a contact interaction is a good approximation at low energies
to a more complete theory described in terms of the exchange of
a heavy charged W-boson, as we mentioned earlier.
Here $U_{KM}$ is a general unitary 3$\times$3 matrix, which is not needed in the
case of the lepton current due to the assumption in the standard
model that the three neutrinos are degenerate.
Consequently Fermi's weak interaction, in its modern guise, contains
an enormous range of physical processes.  In addition in 1972 a
different kind of weak interaction was found, wherein a {\it neutral}
current (which is diagonal in quark and lepton identities)
is coupled to its hermitian conjugate.
\begin{eqnarray}
{\cal J}_\mu&=&{\cal J}_\mu^{\rm had}+{\cal J}_\mu^{\rm lep}=
\left(\psi_e^\dagger\,\,\psi_\mu^\dagger\,\,\psi_\tau^\dagger\right)j'_\mu
\left(\begin{array}{l}
\psi_e\\
\psi_\mu\\
\psi_\tau
\end{array}\right)+\left(\psi_{\nu_e}^\dagger\,\,
\psi_{\nu_\mu}^\dagger\,\,\psi_{\nu_\tau}^\dagger
\right)j'_\mu\left(\begin{array}{l}
\psi_{\nu_e}\\
\psi_{\nu_\mu}\\
\psi_{\nu_\tau}
\end{array}\right)\nonumber\\
&+&\left(\psi_u^\dagger\,\,\psi_c^\dagger\,\,\psi_t^\dagger\right)j'_\mu
\left(\begin{array}{l}
\psi_u\\
\psi_c\\
\psi_t
\end{array}\right)+\left(\psi_d^\dagger\,\,\psi_s^\dagger\,\,\psi_b^\dagger\right)
j'_\mu\left(\begin{array}{l}
\psi_d\\
\psi_s\\
\psi_b
\end{array}\right)
\end{eqnarray}
with
\begin{equation}
{\cal H}_w\sim{G_F\over \sqrt{2}}{\cal J}_\mu^\dagger{\cal J}^\mu
\end{equation}
In this case the interaction arises from the exchange of a heavy
neutral particle---the Z-boson with mass $m_Z\sim 91$ GeV---and can
again be taken to be of contact form for low-energy reactions.

Before leaving this historical journey it is useful to remark on one
additional feature of the weak interaction important to
modern studies---the handedness.  As both quarks and leptons are
spin-1/2 objects, they can be described by four-component Dirac fields
and their currents can be expressed as bilinear forms of such fields
connected by 4$\times$4 Dirac matrices.  Experimentally one finds that the
proper combination is an equal mixture of polar and axial vector
structures
\begin{equation}
\psi^\dagger_aj_\mu\psi_b\equiv\psi_a^\dagger\gamma_0\gamma_\mu(1-\gamma_5)\psi_b
\end{equation}
The Dirac matrix
\begin{equation}
1-\gamma_5=\left(\begin{array}{cc}
~1&-1\\
-1&~1
\end{array}
\right)
\end{equation}
is called a ``chirality'' operator and, for
particles of zero mass, projects out only ``left-handed''
particles---{\it i.e.}, those whose spins are aligned opposite to their
momenta.  In order to see this, consider the Dirac equation for a free
particle of mass $m$
\begin{equation}
(i\not\!{\partial}-m)\psi(x)=0
\end{equation}
where $\not\!{\partial} = \!{\partial}_0 \gamma_0 - \vec{\!{\partial}}
\cdot \vec{\gamma}$.
We use the standard representation for the 4$\times$4 Dirac
matrices\cite{bj}
\begin{equation}
\gamma_0=\left(
\begin{array}{ll}
1&~0\\
0&-1
\end{array}
\right)\quad
\vec{\gamma}=\left(
\begin{array}{cc}
~0&\vec{\sigma}\\
-\vec{\sigma}&0
\end{array}
\right)\label{eq:cc}
\end{equation}
The positive energy plane wave solutions of Eq. \ref{eq:cc} are well-known
\begin{equation}
\psi(x)=\sqrt{E+m\over 2E}\left(\begin{array}{c}
\chi\\
{\vec{\sigma}\cdot\vec{p}\over E+m}\chi
\end{array}
\right)\exp(-ip\cdot x)
\end{equation}
where $\chi$ is a two-component Pauli spinor.
Then in the limit as $m\rightarrow 0$ and E$\rightarrow |\vec{p}|$ this
becomes
\begin{equation}
\psi(x) \begin{array}[t]{c} \longrightarrow \\ {\footnotesize m \rightarrow 0 } \end{array}
\psi_0(x)={1\over \sqrt{2}}\left(
\begin{array}{c}
\chi\\
\vec{\sigma}\cdot\hat{p}\chi
\end{array}
\right)\exp(-ip\cdot x)
\end{equation}
so that
\begin{equation}
(1-\gamma_5)\psi_0(x)=\left\{\begin{array}{cc}
\psi_0(x) & \vec{\sigma}\cdot\hat{p}\chi=-\chi\\
0 & \vec{\sigma}\cdot\hat{p}\chi=\chi
\end{array}\right.
\end{equation}
as claimed.  This result is important, as will be discussed in
the next section, because the neutrino is either massless or {\it extremely}
light.  Therefore, since the neutrino interacts only via the weak
interactions which involve the chirality operator $1-\gamma_5$,
{\it all neutrinos must be left-handed}!\cite{rth}  Similarly, it is easy to
see that
all antineutrinos must be right-handed.

Another way of stating this result is to say that, although Dirac spinors are
{\it four}-component objects, those describing zero mass neutrinos involve
only two of the four components.  This ``two-component neutrino''
theory has been tested in a direct measurement of neutrino helicity in
the reaction\cite{neu}
\begin{equation}
{}^{152}{\rm Eu}(0^-)+e^-\rightarrow{}^{152}{\rm Sm}^*(1^-)+
\nu_e\rightarrow{}^{152}{\rm Sm}(0^+)+\gamma+\nu_e
\end{equation}
The clever idea behind this scheme is that one can select those $\gamma$'s from
the decay
of the Sm excited state which travel oppositely to the direction of the
electron-capture $\nu_e$'s ({\it i.e.} in the
direction of the nuclear recoil) by having them resonantly scatter from
a Sm target.  By angular momentum conservation the helicity of the
downward-going $\gamma$ is the same as that of the upward-traveling
$\nu_e$.  The results of the experiment strongly confirmed the
two-component hypothesis.

It is interesting to note that the chirality structure of the weak current also
explains why the decay of the charged pion proceeds predominantly via
$\pi^+\rightarrow\mu^++\nu_\mu$ rather
than by the mode $\pi^+\rightarrow e^++\nu_e$ which is strongly favored by
phase
space.  The point is that if
the positron were massless, it too would be described by a
two-component theory and any such particle coupled to the weak
interaction would have to be purely right-handed.  Then, as diagrammed in
Figure 3,
the decay of a pion into a positron and a neutrino must be forbidden
because angular momentum conservation prohibits the coupling of
a right-handed positron and left-handed neutrino to a spinless system.  Of
course, in the real world the positron is light but not massless.
Thus the positron decay of the charged pion is not completely forbidden but
rather highly suppressed compared to its muonic counterpart---
\begin{equation}
R_e\equiv{\Gamma(\pi^+\rightarrow e^++\nu_e)\over \Gamma(\pi^+\rightarrow
\mu^++\nu_\mu)}=\left({m_e\over m_\mu}\right)^2\left(m_\pi^2-m_e^2\over m_\pi^2
-m_\mu^2\right)^2=1.23\times 10^{-4}
\end{equation}
which is confirmed by experiment\cite{pdb}
\begin{equation}
R_e^{exp}=(1.230\pm 0.004)\times 10^{-4}
\end{equation}

\begin{figure}
\psfig{bbllx=2.0cm,bblly=3.5cm,bburx=15cm,bbury=8.3cm,figure=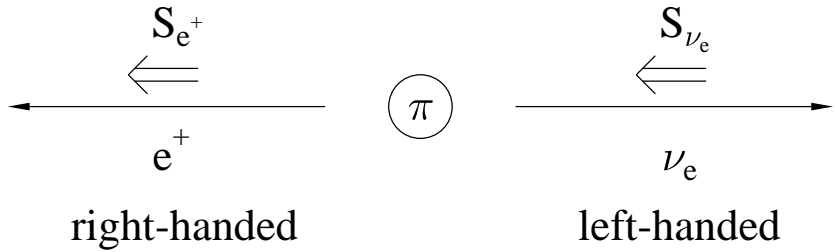,height=1.6in}
\caption{Schematic representation of a pion at rest decaying
to a massless positron and a neutrino.  Such a decay is forbidden
by angular momentum conservation.}
\end{figure}

It is interesting to note one final point about neutrinos which has a
close connection to helicity.  The reader will have noticed
that we have throughout distinguished the $\nu_e$ produced
when a proton beta decays in a nucleus from the $\bar{\nu}_e$
produced in neutron beta decay.  The concept of a distinct
antiparticle is certainly clear for charged leptons like the
electron, as its antiparticle---the positron---carries the opposite
charge.  More generally, particle-antiparticle
conjugation reverses the signs of {\it all} of a particle's additively
conserved
quantum numbers.  The neutrino is
immediately seen to be quite interesting then, as it lacks a
charge, magnetic moment, or other measured quantum number
that would necessarily reverse under such an operation---it
is unique among the leptons
and quarks in that the existence of a {\it distinct} antiparticle
is an open question.

Early on, before the handedness of the weak interaction was
discovered, there appeared to be a simple test of the
particle-antiparticle properties of the neutrino.  If one
defines the $\nu_e$ as the neutrino produced when a proton
decays in a $\beta^+$ source, then one finds that
$\nu_e$'s produce electrons by the reaction
\begin{equation}
\nu_e + n \rightarrow p + e^- \label{eq:aaa}
\end{equation}
but {\it not} positrons in the analogous reaction
\begin{equation}
\nu_e + p \not\rightarrow n + e^+ \label{eq:bbb}
\end{equation}
Similarly if we define the $\bar{\nu}_e$ as the particle
produced in the $\beta^-$ decay of the neutron decay,
then $\bar{\nu}_e$'s produce positrons by the reaction
\begin{equation}
\bar{\nu}_e + p \rightarrow n + e^+ \label{eq:ccc}
\end{equation}
but not electrons by the reaction
\begin{equation}
\bar{\nu}_e + n \not\rightarrow p + e^- \label{eq:ddd}
\end{equation}
Thus it would appear that the $\nu_e$ and the $\bar{\nu}_e$ are operationally
distinct.
In fact, the absence of the reactions in Eqs. \ref{eq:bbb},\ref{eq:ddd}
became apparent around 1950 from an experiment done by nature,
a form of natural radioactivity known as double
beta decay.  If, for example, the reaction in Eq. \ref{eq:ddd}
were allowed, certain nuclei could undergo the second-order
weak decay
\begin{equation}
(A,Z) \rightarrow (A,Z+1) + e^- + \bar{\nu}_e \rightarrow (A,Z+2) +
2e^- \label{eq:eee}
\end{equation}
where the neutrino produced in the first decay is reabsorbed
by the nucleus, producing a final state with two electrons
and no neutrinos.  The absence of such ``neutrinoless double beta
decay,"
which has a distinctive experimental signal because the entire
energy release is carried off by the electrons,
thus seemed to show that the $\nu_e$ and $\bar{\nu}_e$ were
indeed distinct particles \cite{haxbb}.  This prompted the introduction of
a distinguishing quantum number, lepton number.  The $\nu_e$
and electron were assigned $l_e$ = +1, the $\bar{\nu}_e$ and
positron $l_e$ = -1.  The assumption of an additively
conserved lepton number in weak interactions then allows
the reactions in Eqs. \ref{eq:aaa},\ref{eq:ccc}, but explains the
absence of the reactions in Eqs. \ref{eq:bbb},\ref{eq:ddd},\ref{eq:eee}.
A neutrino with a distinct antineutrino is called a Dirac
neutrino.

However the discovery of the apparent exact handedness of the
weak interaction invalidates this simple conclusion.
All of the results are also explained by the assignments
\begin{equation}
\nu_e \rightarrow \nu_e^{LH}~~~\mathrm{and}~~~\bar{\nu}_e
\rightarrow \nu_e^{RH}
\end{equation}
and a weak interaction that violates parity maximally.
Here $RH$ denotes a right-handed particle and $LH$ a left-handed one.
Thus the possibility that the neutrino is its own antiparticle ---
a so-called Majorana neutrino --- is still open.
In this case a reaction like that of Eq. \ref{eq:eee} is not
forbidden by an exact additive conservation law, but rather by {\it helicity}.
Therefore if a Majorana neutrino had a small mass, neutrinoless
double beta decay would occur, but the decay rate would
by suppressed by the small quantity
\begin{equation}
\left( {m_\nu \over E_\nu} \right)^2
\end{equation}
where $E_\nu \sim$ 50 MeV is an energy characteristic of the virtual
neutrino emitted and reabsorbed in the decay.  Modern searches for
neutrinoless double beta have established limits on
half lives of $\sim 10^{25}$ y, corresponding to a Majorana
neutrino mass below 1 eV \cite{klap}.

Given that the familiar charged leptons have only Dirac masses, it
is natural to ask why neutrinos, which can have two kinds
of masses, would then be the only massless leptons in the standard model.
The absence of Dirac neutrino masses in the standard model
follows from the need to have both left-hand and right-handed
fields in order to construct such masses.  We have noted that neutrinos
interact only weakly and that weak interactions involve only
left-handed components of the fields.  The standard model,
being very economical, has no right-handed neutrino fields
and thus no Dirac neutrino masses.  However, the absence of Majorana masses
has a more subtle explanation.  One can construct a left-handed
Majorana mass with the available standard model neutrino
fields, but it turns out this term is not ``renormalizable,"
{\it i.e.}, it generates infinities in the theory.  Our point-like
Fermi $\beta$ decay theory is another example of a
nonrenormalizable theory, though it works quite well in the
domain of low-energy weak interactions.  If we were to
relegate the standard electroweak model to a similar status ---
that of an effective theory --- Majorana mass terms could then
be introduced.  In effect, most extensions of the standard model
do precisely that, and also generally introduce new fields
such as those creating right-handed neutrinos.  Thus almost
all theorists, believing the standard model is incomplete and
must be extended in some such ways, also believe that neutrinos
have masses.  Indeed, the puzzle is rather to explain why these
masses are so much smaller than those of charged particles.

With this historical background out of the way, we now move to
consider aspects of the neutrino that have recently been in
the news --- masses and mixings.

\section{Neutrino Mass: Direct Measurements}

The issue of whether the neutrino has a nonzero mass has been long one
of interest.  That any such mass must be small could be seen from the feature
that
the maximum energy measured in the beta spectrum agreed to high
precision with the mass difference of initial and final nuclear
states.  However, Fermi, in his seminal paper on beta decay, noted that
this question could be answered more definitively by carefully
studying the endpoint of the electron spectrum---it is possible
to plot the spectrum in such a way that a nonzero mass would
be revealed as a distortion at the endpoint tangent to
(perpendicular to ) the energy axis, as shown in Figure 4.
In the years since Fermi's paper there has been a series of
such measurements, with steadily
increasing precision.  It is clear that use of a
$\beta$ decay parent nucleus with a relatively low
Q-value is helpful, as a larger fraction of total decays then
resides within a given interval from the endpoint.  Most experimenters have
selected
tritium, which has an 18.6 keV endpoint.
An early tritium measurement by Hamilton, Alford, and Gross found an
upper limit $m(\bar{\nu}_e) \lsim 250$ keV\cite{hag}.  A few years later,
Bergkvist, by
combining electrostatic and magnetic spectrometric methods, was able to
reduce the limit substantially---$m(\bar{\nu}_e) \lsim
60$ eV \cite{kb}.  Then in 1980 Lubimov et
al., using a high-precision toroidal spectrometer and tritium in the form of the
valine molecule (${\rm C}_5{\rm H}_{11}{\rm NO}_2$),
claimed the first nonzero mass---$14$ eV$ \lsim  m(\bar{\nu}_e)
\lsim  46$ eV,\cite{lub}, a result that set
off a flurry of new, high precision experiments.
Before discussing the results, however, we first examine one of the reasons
these experiments are important---the cosmological significance
of a massive neutrino.

\begin{figure}
\psfig{bbllx=-0.5cm,bblly=10.cm,bburx=14cm,bbury=20.5cm,figure=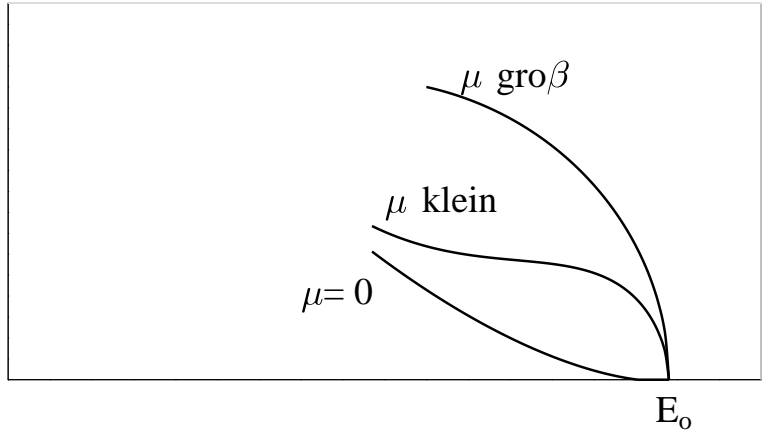,height=2.3in}
\caption{A reproduction of curves from Fermi's paper (Ref. \cite{fermi}) showing the
shape of the $\beta$ spectrum near the endpoint for the cases of
zero and nonzero neutrino masses.}
\end{figure}

We are all aware that at the present time the universe is expanding.
However, what will be its ultimate fate?  Will it continue to expand
forever, or will the expansion slow and finally reverse?  In order to see what
role neutrinos may play in answering this question, we explore their effects in
an
expanding homogeneous and isotropic universe.  Consider a small test mass
$m$ which sits on the surface of a spherical chunk of this
universe having radius R.  If the mean energy density of the universe
is $\rho$, then the mass contained inside the spherical volume is
\begin{equation}
M(R)={4\over 3}\pi R^3\rho
\end{equation}
The potential energy of the test mass, as seen by an observer at the
center of the sphere, is
\begin{equation}
U=-G{M(R)m\over R}
\end{equation}
while its kinetic energy is
\begin{equation}
T={1\over 2}mv^2={1\over 2}m\left({dR\over dt}\right)^2
\end{equation}
By Hubble's Law the expansion velocity is given by
\begin{equation}
v=HR
\end{equation}
where $H={1\over R}{dR\over dt}$ is the Hubble constant.  Although the
size of $H$ is still uncertain, most values are in the range
\begin{equation}
50 {\rm km/sec/Megaparsec} \lsim H_{\rm exp}\lsim 100 {\rm km/sec/Megaparsec}
\end{equation}
(with perhaps $\sim$ 65 km/sec/Megaparsec being the best value).
The total energy of the test particle is then
\begin{equation}
E_{\rm tot}=T+U={1\over 2}mR^2(H^2-{8\over 3}\pi\rho G)
\end{equation}
and the fate of the universe depends on the sign of this number, or
equivalently
with the relation of the density to a critical value
\begin{equation}
\rho_c={3H^2\over 8\pi G}\sim 2\times 10^{-29}{\rm g/cm}^3,
\end{equation}
{\it i.e.}
\begin{equation}
\begin{array}{cc}
\rho<\rho_c& \Rightarrow \mbox{continued expansion}\\
\rho>\rho_c& \Rightarrow \mbox{ultimate contraction}
\end{array}
\end{equation}
It is not yet clear which situation describes
our universe.  Analysis of the dynamics of gravitationally bound systems
via the virial theorem and comparison with measured luminosities yields
the ``visible'' mass density
\begin{equation}
\rho_{\rm vis}\simeq 0.02\rho_c
\end{equation}
However, Doppler studies of the rotation rates of
spiral galaxies indicate that these systems are
much more massive than their luminosities seem to suggest
\begin{equation}
\rho_{\rm rot}\sim 20\rho_{\rm vis}
\end{equation}
The origin of the ``dark matter'' responsible for this discrepancy
is a matter of current study: there are several possibilities, including
massive
neutrinos.  But regardless of the origin of the dark matter, it
appears that the energy density of our universe is within an order of
magnitude of the critical value required for closure.

Before discussing the relevance of neutrinos to the dark matter, we
consider the more familiar case of the relic photon spectrum, which
has been carefully studied recently via the COBE satellite.
In the early universe electromagnetic interactions such as
\begin{equation}
e^++e^-\leftrightarrow \gamma +\gamma ,
\qquad\gamma+e^-\leftrightarrow\gamma+e^-
\end{equation}
kept the photons in thermal equilibrium with the electrons and nucleons.
However once the universe cooled sufficiently---to $T \sim$ 1 eV at
about $10^5$ years after the big bang---electrons and protons combined to
form neutral hydrogen.  The absence of charged particles
rendered the universe transparent to photons, which then thermally
decoupled from the rest of the matter.  At the time of
``recombination"---the epoch when neutral hydrogen is formed---the
photons were characterized by a blackbody spectrum at
a temperature $T_\gamma^{\rm recomb}\sim$ 4000K.  Because of the subsequent
expansion of the universe, the spectrum today has been redshifted
to a temperature $T_\gamma^{\rm now} \sim 2.72$K, as was established in
the COBE measurements.  The energy density of these relic
photons is given by Stefan's Law
\begin{equation}
\rho_\gamma=2\int{d^3q\over (2\pi)^3}{q\over \exp(q/T)-1}=4\sigma
T_\gamma^4\sim 4\times 10^{-34}\,{\rm g/cm}^3
\end{equation}
where $\sigma = \pi^2/60$ is the Stefan-Boltzmann constant.

An analogous relic neutrino spectrum must exist from the early
universe.  Such neutrinos were originally kept in thermal equilibrium
with electrons and other matter by the weak interactions
\begin{equation}
e^++e^-\leftrightarrow \nu_e+\bar{\nu}_e,\quad e^\pm+\nu_e\leftrightarrow
e^\pm+\nu_e,\quad e^\pm+\bar{\nu}_e\leftrightarrow e^\pm+\bar{\nu}_e
\end{equation}
However, once the temperature cooled to about $10^{10}$K the
weak reaction rates---which depend on $\sigma_{\rm weak} n v$, where
$n$ is the lepton density and $v$ the relative velocity, and vary as $T^5$---can no longer keep
up with the expansion rate,
$H \sim \dot{R}/R \sim T^2$, which slows less rapidly.
The neutrinos then drop out of equilibrium with the charged leptons.  From this
point they are essentially decoupled from the rest of the universe,
but, of course, cool as expansion proceeds.  The neutrino energy density at
the present time is given by
\begin{equation}
\rho_\nu={7\over 2}N_\nu \sigma T_\nu^4
\end{equation}
where $N_\nu=3$ is the number of ``massless'' ($m_\nu<T$) two-component
neutrino generations\cite{con}.
Note, however, that the present neutrino and photon temperatures are
different: while above $\sim 10^{10}$K the reaction
$e^++e^-\leftrightarrow\gamma+\gamma$ proceeds in both directions,
below this temperature the photon energy is no longer sufficient to
produce pairs.  Thus subsequent reactions proceed only to the right, resulting
in a reheating process that raises the photon temperature with respect
to the neutrinos by the factor
\begin{equation}
\left({\rho_\gamma+\rho_{e^-}+\rho_{e^+}\over
\rho_\gamma}\right)^{1\over 3}=\left({11\over 4}\right)^{1\over 3}
\end{equation}
The present-day relic neutrino temperature is thus
\begin{equation}
T_\nu=T_\gamma\left({4\over 11}\right)^{1\over 3}=1.94K
\end{equation}
The corresponding average neutrino energy is only $\sim 10^{-3}$ eV, so
that scattering cross sections are unobservably small.  But such relic neutrinos
do contribute to the mass density of the universe, yielding in
the case of massless neutrinos
\begin{equation}
\rho_\nu=\rho_\gamma\times {7\over 8}\times N_\nu\times\left({4\over
11}\right)^{4\over 3}\simeq 0.7\rho_\gamma
\end{equation}
If, however, neutrinos have a nonzero mass, then a contribution comparable
to the critical density results if the sum of the masses for the three
neutrino species is as little as $\sim$ 25 eV.  Thus neutrino masses easily
compatible with existing experimental limits could close the
universe and ultimately lead it to recollapse.  In particular,
since this number is within the range found in ref. \cite{lub} for the electron
antineutrino {\it alone}, confirmation of that experiment was clearly
crucial to cosmology.

Stimulated by the Lubimov result, several groups attempted
improved versions of this experiment.  One of the criticisms of the Russian
experiment was the use a tritiated valine source, which introduced
a substantial uncertainty because of binding effects and because
the contribution of molecular excited states, populated in the
beta decay, to the energy loss could not be calculated easily.
Thus a group at Los Alamos used a much simpler source,
gaseous tritium molecules, tackling at the
same time the serious safety issues associated with handling a
kilocurie of this gas.  After a series of measurements with
a carefully constructed magnetic spectrometer that filled an entire room,
they found $m_{\bar{\nu}_e} \lsim 9.3$ eV \cite{la}.  An experiment
at Mainz using a frozen tritium source reported a similar limit,
$m_{\bar{\nu}_e} \lsim 7.2$ eV \cite{mai}, and a Livermore group using
gaseous tritium and a toroidal magnetic spectrometer
achieved comparable statistics \cite{lll}.  All of these experiments,
however, were troubled by a puzzling excess of events near
the endpoint.  Although this problem has been described
as a {\it negative} value for $m_{\bar{\nu}_e}^2$ (!), in fact
this was too simple a characterization: each of the groups
described the anomaly with a different functional form.  While
standard statistical techniques were then used to establish
the above bounds on $m_{\bar{\nu}_e}^2$, it is apparent than an
unknown systematic contributing excess events at the endpoint
could also mask the effects of a {\it positive} $m_{\bar{\nu}_e}^2$.
This has tended to weaken the community's confidence in the stated bounds.
Ongoing experiments at Mainz and
Troitsk now claim limits of $\sim$ (3-5) eV and, while
improved resolution and understanding of energy loss in the
target has significantly mitigated the negative
$m_{\bar{\nu}_e}^2$ problem, certain anomalies remain in
the endpoint region.

Because of this situation, it was fortunate that a special
event occurred that established an independent bound on the
$\bar{\nu}_e$ mass.  This was
the observation of Supernova 1987A, which was found in the
southern hemisphere on February 23rd of that year---{\it cf.} Figure 5.
SN1987A resulted from the explosion of a star in the Large Magellanic
Cloud about 170,000 years ago, the light (and neutrinos) from which
finally reached
earth in 1987.  This was the first such optical supernova in our vicinity in
nearly four hundred years, the previous occurrence having been noted
by Kepler in 1604!

\begin{figure}
\psfig{bbllx=0.5cm,bblly=4.0cm,bburx=18cm,bbury=18.5cm,figure=barry_fig5.ps,height=2.4in}
\caption{Views of a region in the Large Magellanic Cloud before (right)
and after (left) the morning of February 27, 1987: SN1987A is clearly
visible.  Copyright Anglo-Australian Observatory.
Photograph by David Malin (http://www.aao.gov.au/local/www/dfm/aat050.html).}
\end{figure}

A brief discussion of stellar evolution is needed in order to
make the connection between SN1987A and neutrinos.
Small stars like our sun live relatively quiescent lives,
spending billions of years slowly burning hydrogen to helium in their
hot dense cores, with the liberated energy maintaining
the electron gas pressure that stabilizes the star against
gravitational collapse.  A Type II supernova - the type to
which SN1987A belongs - is the last evolutionary stage of
a more massive star, in excess of 10 solar masses.  Like our
sun, such a star begins its lifetime burning
the hydrogen in its core under the conditions of hydrostatic
equilibrium.  When the hydrogen is exhausted, the core contracts
until the density and temperature are reached where
helium can ignite via the 3$\alpha \rightarrow
^{12}$C reaction.  The He is then burned to exhaustion.
This pattern (fuel exhaustion, contraction, and ignition of the
ashes of the previous burning cycle) repeats several times,
leading finally to the explosive burning of $^{28}$Si to Fe.
For a heavy star, the evolution is rapid: the star has to work
harder to maintain itself against its own gravity, and therefore
consumes its fuel faster.  A 25 solar mass star would go through
all of these cycles in about 7 My, with the final explosive Si
burning stage taking only a few days!

Iron is the most strongly bound nucleus in the periodic table.
Thus once the Si burns to produce Fe, there is no further source
of nuclear energy with which to support the star.  So, as the last
remnants of nuclear burning take place, the core is largely
supported by the electron degeneracy pressure.
When enough ash accummulates so that the iron core exceeds
the Chandrasekhar limit \cite{ch}---a limit of about 1.4
solar masses above which it is no longer stable--the core
begins to collapse.  Gravity
does work on the infalling matter, leading to rapid heating
and compression of the iron, and ultimately ``boiling off" $\alpha$'s
and a few nucleons from the nuclei.  At the same time, the
electron chemical potential is
increasing, making electron capture on nuclei and any free
protons favorable,
\begin{equation}
 e^- + p \rightarrow \nu_e + n.
\end{equation}
Both the electron capture and
the nuclear excitation and disassociation take energy out of the electron gas,
which is the star's only source of support.  This means that
the collapse is very rapid.  Indeed, numerical simulations find that
the iron core of the star ($\sim$ 1.2-1.5 solar masses) collapses
at about 0.6 of the free fall velocity.

In the early stages of the infall the $\nu_e$'s readily escape.
But neutrinos become trapped when a
density of $\sim$ 10$^{12}$g/cm$^3$ is reached, at which point they
begin to scatter off the matter through
both charged current and coherent neutral current processes.  The
neutral current neutrino scattering off nuclei is particularly
important, as the scattering cross section involves the total nuclear
weak charge, which is approximately the
neutron number.  This process transfers very little energy because
the mass energy of the nucleus is so much greater than the
typical energy of the neutrinos.  But momentum is exchanged.
Thus the neutrino ``random walks" out of the star,
frequently changing directions.  When the
neutrino mean free path becomes sufficiently short, the ``trapping
time" of the neutrino begins to exceed the time scale for the
collapse to be completed.  This occurs at a density of about
10$^{12}$ g/cm$^3$, or somewhat less than 1\% of nuclear density.
After this point, the energy released by further gravitational
collapse is trapped within the star.
If we take a neutron star of 1.4 solar masses and a radius of
10 km, an estimate of this gravitational energy is
\begin{equation}
 {G M^2 \over 2R} \sim 2.5 \times 10^{53} \mathrm{ergs}.
\end{equation}

The collapse continues until nuclear
densities are reached.  As nuclear matter is rather
incompressible (compression modulus $\sim$ 300 MeV),
the nuclear equation of state then halts the collapse:
maximum densities of 3-4 times nuclear density are reached, {\it e.g.},
perhaps $6 \cdot 10^{14}$ g/cm$^3$.
This sudden braking of the collapse generates a series of pressure
waves which travel outward through the iron core and
collect at the sonic point (where the infall velocity of
the iron matches the sound speed in iron), eventually forming
a shock wave that travels out through the mantle of the star.
Theory suggests that this shock wave in combination neutrino
heating of the matter is responsible for the ejection of the mantle---and
thus the spectacular optical display we saw in 1987.

Yet this visible display represents less than 1\% of the energy locked
inside the protoneutron star.  After the core collapse and
shock wave formation are completed---events that take on the
order of 10's of milliseconds---the protoneutron star cools
by emitting a largely invisible radiation, a huge fluence of
neutrinos of all flavors.  Essentially all
of the 3 $\cdot 10^{53}$ ergs released in the collapse is
radiated via electron, muon, and tauon neutrinos over the next 10 seconds.

In 1987 there existed deep underground two detectors consisting
of tanks of ultrapure water instrumented with phototubes.
They were originally constructed for the purpose of
detecting proton decay, {\it e.g.} $p\rightarrow e^++\gamma$,
a process predicted
by certain grand unified theories of elementary particle structure,
but not yet observed.
The $\gamma$ ray and
Cerenkov light generated by the passage
of the positron through the water would be detected by the
phototubes, signaling a proton decay event.
One detector was located in a Morton
salt mine outside Cleveland, Ohio, while the second was on the other
side of the globe in the Kamioka mine within the Japanese alps.
While the search for a proton decay signal proved futile,
on February 23, 1987 both detectors recorded
a handful of events associated with the passage
of a strong burst of antineutrinos through the tanks,
a few of which initiated the reaction $\bar{\nu}_e+p\rightarrow e^++n$.  Within
several hours a Canadian astronomer observing in Chile noted a
new bright star.  This was SN1987A.

How do these supernova neutrino events limit the neutrino mass?  If a neutrino
has
mass, neutrinos of different energy will travel with different
velocities according to
\begin{equation}
v={p_\nu\over \sqrt{p_\nu^2+m_\nu^2}}\simeq 1-{m_\nu^2\over 2E_\nu^2}
\end{equation}
rather than with the uniform velocity $c$.
Consequently the time of arrival of higher energy neutrinos will be
earlier than their lower energy counterparts. The difference in
arrival times is then
\begin{equation}
{\delta t\over t}\sim{\delta v\over v}\sim {m_\nu^2\over
E_\nu^2}{\delta E_\nu\over E_\nu}
\end{equation}
where $t$ is the total time in transit from the Large Magellanic Cloud.
The eleven events seen by the Kamiokande detector and the eight seen
in Ohio have an energy spread $\delta E_\nu\sim 10$ MeV
and an average energy $E_\nu$ 2-3 times larger.
The events arrived
over a ten second time span.  Thus we find in a simple
analysis
\begin{equation}
m_{\bar{\nu}_e}\lsim E_\nu\left({\delta t\over t}{E_\nu\over \delta
E_\nu}\right)^{1\over 2}\sim 10\,{\rm MeV}\left({10 {\rm sec}\over
10^{13} {\rm sec}}\right)^{1\over 2}\sim 10\, {\rm eV}
\end{equation}
A more careful analysis, which takes into account the time and energy
distribution, yields a similar limit---$m_{\bar{\nu}_e}\lsim 20$ eV.
It is remarkable that a handful of neutrino events from a star that
lived and died long before the advent of civilization
places a bound on the neutrino mass comparable to that gained from many
years of high precision weak interactions studies!

We have focused on electron neutrino mass measurements because
these have achieved the greatest precision.
Bounds have also been established on the masses of the $\nu_\mu$
and $\nu_\tau$
\begin{equation}
m_{\nu_\mu}\lsim 170\, {\rm keV},\qquad m_{\nu_\tau}\lsim 24 \,{\rm MeV}
\end{equation}
from careful kinematic analyses of decays
such as $\pi^+\rightarrow\mu^++\nu_\mu$ and
$\tau\rightarrow 5\pi+\nu_\tau$.
The short lifetimes of the $\pi$ and $\tau$ serve to limit
the precision of the bounds.  It is worth noting that some
theories of neutrino mass predict that the $\nu_\tau$ will be the
heaviest neutrino; {\it e.g.}, in some models neutrino masses
scale as the squares of the masses of the corresponding
charged leptons.  From this perspective the limits directly above are not
necessarily
less significant than the tighter bounds established on the
$\bar{\nu}_e$.

This discussion can be succinctly summarized:
there exists no evidence for massive neutrinos from
direct measurements.  Naturally, when increasingly precise experiments
continue to yield values consistent with zero, it is tempting to
assume that the number really {\it is} zero.  Indeed, as mentioned above,
in the so-called
``standard model'' of elementary particles, the neutrino mass is
assumed to vanish.  This does not follow from any fundamental
principle, however, and could be modified accordingly if the evidence
were to change, as we have noted before.
That is precisely what has been happening over the
last several years, as we will now describe.

\section{Probing Masses through Neutrino Mixing}

Significant improvements in the ``direct'' neutrino mass measurements
described above will probably require considerable time and effort.
For this reason there is great interest in measurements of
a different type---those exploiting neutrino mixing---which
might be able to probe far smaller masses, albeit in a less
direct fashion.  The essential idea can be traced to the seminal
paper of Pontecorvo \cite{pont}, who first pointed out that neutrino
oscillations would occur if the neutrino states of definite
mass do not coincide with the weak interaction eigenstates.
To understand neutrino oscillations it is helpful to first
consider the more familiar phenomenon of Faraday rotation,
the rotation around the beam direction of the polarization
vector of linearly polarized light as the light propagates
through a magnetized material.  Faraday rotation occurs
because the index of refraction (and thereby the potential acting on the light)
depends on the state of circular polarization---{\it i.e.},
the two states of definite circular polarization
($|\chi_\pm>=\sqrt{1\over 2}(|+x>\pm i|+y>)$
evolve in time with distinct phases.
Thus if $|\chi(t=0)> = |+x> = \sqrt{1\over 2} (|\chi_+>+|\chi_->)$,
then
\begin{eqnarray}
|\chi(t)>&=&\sqrt{1\over 2}\left(|\chi_+>\exp(-i{\omega_0\over
n_+}t)+|\chi_->\exp(-i{\omega_0\over n_-}t)\right)\nonumber\\
&=&\exp(-i{\omega_0\over 2}{n_++n_-\over
n_+n_-}t) \left( |+x>\cos({\omega_0\over 2}{n_+-n_-\over
n_+n_-}t) \right. \nonumber\\
&-& \left. |+y>\sin({\omega_0\over 2}{n_+-n_-\over n_+n_-}t)\right)
\end{eqnarray}
so that the polarization vector rotates with frequency
\begin{equation}
\omega_{pol} = {\omega_0 \over 2} {n_+ - n_- \over n_+n_-}
\end{equation}

Although we have three neutrino families, the essential physics of
oscillations is illustrated very well by considering the
interactions of only two of these, which we choose to be
the $\nu_e$ and $\nu_\mu$.  In this limit the situation is similar to the
two-state
Faraday rotation problem discussed above.  The relevant part of the
weak current involves the combination
$\bar{e}\nu_e+\bar{\mu}\nu_\mu$ and is given by Eq. \ref{eq:dd}.  This
effectively defines the $\nu_e$ and $\nu_\mu$: they are the
neutrino weak interaction eigenstates, the neutrinos accompanying
the electron and muon, respectively, when these particles are
weakly produced.  Yet there is a second Hamiltonian, the free
Hamiltonian describing the propagation of an isolated neutrino.
The eigenstates of this Hamiltonian are the mass eigenstates.
If the two mass eigenstates are distinct (and thus at least one
is nonzero), then in general the eigenstates diagonalizing the mass
Hamiltonian will {\it not} diagonalize the weak interaction.  If we label the
mass eigenstates as $|\nu_1>$ and $|\nu_2>$, then
\begin{eqnarray}
|\nu_1>&=&\cos\theta|\nu_e>+\sin\theta|\nu_\mu>\quad \mbox{with mass }
m_1\nonumber\\
|\nu_2>&=&-\sin\theta|\nu_e>+\cos\theta|\nu_\mu>\quad \mbox{with mass }
m_2\label{eq:ee}
\end{eqnarray}
where $\theta$ is a mixing angle which distinguishes the mass and
weak eigenstates.  Also suppose that at time $t=0$ an electron
neutrino is produced with fixed momentum $\vec{p}$
\begin{equation}
|\psi(t=0)>=|\nu_e>=\cos\theta|\nu_1>-\sin\theta|\nu_2>
\end{equation}
The mass eigenstates propagate with simple phases, as
they are the eigenstates of the free Hamiltonian.  At a
distance $\sim ct$ from the source the neutrino state is
\begin{equation}
|\psi(t)>=\cos\theta|\nu_1>e^{-iE_1t}-\sin\theta|\nu_2>e^{-iE_2t}
\end{equation}
where $E_i=\sqrt{p^2+m_i^2}$.  Projecting back upon weak
eigenstates, we have
\begin{eqnarray}
<\nu_e|\psi(t)>&=&\cos^2\theta e^{-iE_1t}+\sin^2\theta
e^{-iE_2t}\nonumber\\
<\nu_\mu|\psi(t)>&=&\cos\theta\sin\theta(e^{-iE_1t}-e^{-iE_2t})
\end{eqnarray}
Then, noting that
\begin{eqnarray}
E_1 \sim p + {m_1^2+m_2^2 \over 4p} - {\delta m^2 \over 4p} \nonumber\\
E_2 \sim p + {m_1^2+m_2^2 \over 4p} + {\delta m^2 \over 4p}
\end{eqnarray}
where $\delta m^2 = m_2^2 - m_1^2$, we find at $t>0$ a probability
\begin{equation}
p(t) = |< \nu_\mu | \psi(t) >|^2 = \sin^22\theta \sin^2{\delta m^2 t \over 4p}
\label{eq:osc}
\end{equation}
that the $\nu_e$ will have transformed into a $\nu_\mu$.  This change of
neutrino identity is called a ``neutrino oscillation," due to the
time- or distance-dependent oscillation in $p(t)$.  This phenomenon
is a sensitive test of neutrino masses given nondegenerate neutrinos
and a nonzero mixing angle $\theta$.  As the standard electroweak
model describes neutrinos as massless, the observation of
neutrino oscillations would constitute definitive evidence of
physics beyond the standard model.

Three sources of neutrinos have been exploited by experimentalists
seeking (and apparently finding!) neutrino oscillations:

\subsection{Accelerator and Reactor Experiments}
Such experiments exploit an accelerator or reactor
to produce a large flux of neutrinos/antineutrinos.  Measurements
are then done downstream to determine whether the character of
the neutrinos has changed.  Such experiments fall into one of two
categories.  In the first---called ``disappearance" experiments---one
looks for deviations in the expected flux of neutrinos
of a definite type.  For example, the flux from a reactor is
overwhelmingly of the $\bar{\nu}_e$ type.  Thus oscillations
into a second flavor would lead to an unexpected reduction in
the flux some distance downstream, which might be monitored through
reactions such as $\bar{\nu}_e + p \rightarrow e^+ + n$.
Even if the reactor flux were poorly understood, a definitive
signal could result from detectors placed at different distances
$L_1$ and $L_2$ from the reactor, as shown in Fig. 6a.
In the absence of oscillations the flux must fall off as $1/L^2$
(apart from corrections due to the finite size of the source).
However oscillations superimpose an additional factor of $1-p(t)$,
leading to a modulation that clearly
signals the ``new physics.''  The second type
of oscillation experiment, called an ``appearance'' experiment, is
a search for the product of the oscillation.  For example, at an
accelerator like the Los Alamos Meson Physics Facility (LAMPF),
proton interactions in the beam stop produce large numbers of
$\nu_\mu$'s, $\bar{\nu}_\mu$'s, and $\nu_e$'s, but very few $\bar{\nu}_e$'s:
the reaction chains involve $\pi^+ \rightarrow \mu^+ + \nu_\mu$
followed by $\mu^+ \rightarrow e^+ + \nu_e + \bar{\nu}_\mu$, or
$\pi^- \rightarrow \mu^- + \bar{\nu}_\mu$ followed by
$\mu^- + (A,Z) \rightarrow (A,Z-1) + \nu_\mu$.  It is the
importance of this last reaction
that prevents the $\mu^-$ from ondergoing the free
decay $\mu^- \rightarrow e^- + \bar{\nu}_e + \nu_\mu$---negative
muons are slowed by their interactions with nuclei in
the beam stop and captured into atomic orbitals, then
rapidly cascade electromagnetically into the $1s$ orbital.
There the $\mu^-$ wave function has a strong overlap with the
nucleus (especially for the high Z nuclei in the beam stop),
allowing the $\mu^-$ capture to proceed much more rapidly than
the free decay.  Thus there are very few
electron-type antineutrinos \cite{anti} and
a sensitive search for the flavor oscillation $\bar{\nu}_\mu
\rightarrow \bar{\nu}_e$ can be done by looking for the
appearance of a flux of $\bar{\nu}_e$'s downstream from the
beam stop, as shown in Figure 6b.  Such neutrinos can be
efficiently detected via the reaction $\bar{\nu}_e + p \rightarrow
n + e^+$.  The rate should depend on $1/L^2$ modulated by
$p(t)$.

\begin{figure}
\psfig{bbllx=0.0cm,bblly=2.6cm,bburx=18cm,bbury=11cm,figure=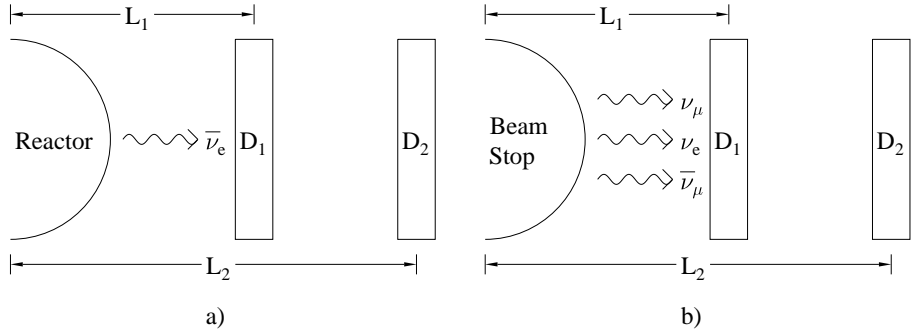,height=2.2in}
\caption{a) A neutrino oscillation experiment of the ``disappearance"
type---a flux of reactor antineutrinos is monitored by detectors
at various distances from the source---in which deviations from a
simple $1/L^2$ falloff are sought;  b) A neutrino oscillation
of the ``appearance" type: detectors are placed at various
distances from a beam stop neutrino source in order to detect
the appearance of an oscillatory $\bar{\nu}_e$ component.}
\end{figure}

The results of such experiments are generally presented in terms of a
two-dimensional diagram with $\delta m^2$ plotted against
$\sin^22\theta$.  Of course, if no signal is observed in such an
experiment it does not necessarily imply that no oscillations are
occurring.  It could be that the mixing angle $\theta$ is very tiny
(and thus $p(t)$ too small to be detected), or that $\delta m^2$ is too small
(so that effects appear only at distances $L$ so large that the neutrino
flux has fallen below detectable limits).
In order to quantify this assertion note that we can write the
oscillation probabilty in terms of path length
\begin{equation}
p(t)\simeq \sin^22\theta\sin^2{\delta m^2L\over 4E_\nu}
\end{equation}
which shows that the maximum effect occurs for
\begin{equation}
{\delta m^2 L \over 4 E_\nu} = {\pi \over 2} \Rightarrow
L = {2 \pi E_\nu \over \delta m^2}
\end{equation}
Thus a smaller $\delta m^2$ implies a longer oscillation length,
so that a more intense neutrino source is required to
combate the associated $1/L^2$ fall off in the flux.
Experiments have ruled out a large portion of the $\sin^22\theta-\delta m^2$
space,
{\it e.g.}, $\sin^22\theta \gsim 0.01$ and $\delta m^2 \gsim 0.1$ eV$^2$,
as shown in Figure 7.  However one experiment---LSND at
Los Alamos---has claimed a positive signal for neutrino oscillations.
This experiment, situated downstream from the LAMPF beamstop,
uses a 52,000 gallon tank of mineral oil and a small amount of liquid
scintillator instrumented with 1220 phototubes.  A neutrino
event $\bar{\nu}_e+p\rightarrow n+e^+$ is indicated by a combination
of Cerenkov and scintillator light produced by the positron followed
(after a couple of hundred microseconds) by a 2.2 MeV gamma ray from
the capture of the produced neutron, $n+p\rightarrow d+\gamma$.
Only 0.0000000001 percent of the LAMPF neutrinos interact in the tank.
Thus the challenge is to distinguish this tiny signal from the
$3 \times 10^8$ cosmic rays which pass through the tank each day.
After very careful numerical simulations, the LSND collaboration
announced they had detected 22 $\bar{\nu}_e$ events compared to
an anticipated background of $4.6 \pm 0.6$ events \cite{lsnd}.
This excess of events is consistent with neutrino oscillations
for the values of $\delta m^2$ and $\sin^22\theta$ shown in
Figure 7: the narrow allowed region includes the ranges
$\delta m^2 \sim 0.2-2$ eV$^2$ and $\sin^22\theta \sim 0.03-0.003$.
Since publication of the first paper, the collaboration's data set
has grown to $\sim$ 60 events.  Concurrently, the competing
KARMEN group of the Rutherford Laboratory in England has found no such oscillation signal, though the sensitivity
achieved to date leaves a substantial portion of the LSND
allowed $\delta m^2 - \sin^22\theta$ region untested.  An improved
experiment at Fermilab, which should have thousands of events if
LSND is correct, has recently been approved and should yield
results by 2002.

\begin{figure}
\psfig{bbllx=-0.5cm,bblly=4.0cm,bburx=17cm,bbury=23.5cm,figure=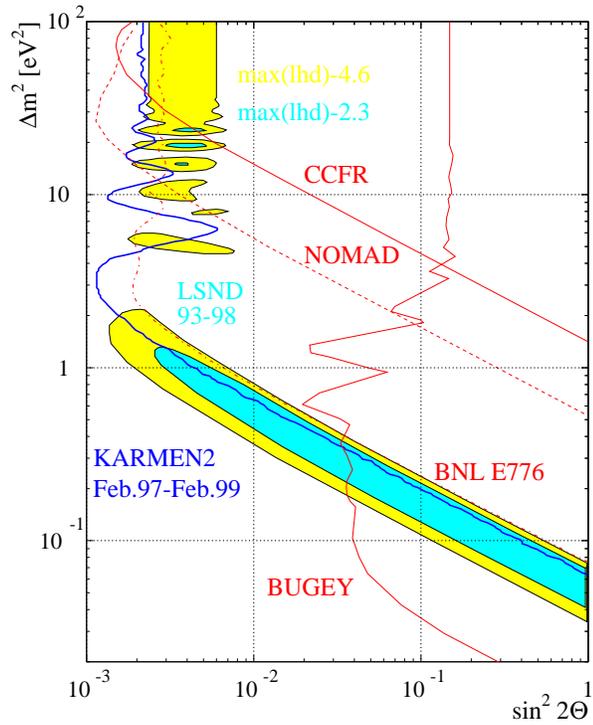,height=3.6in}
\caption{Regions of $\delta m^2$ and $\sin^2 2\theta$ ruled out
by various null experiments are shown along with the allowed
region corresponding to the LSND results.  Thus the portion of
the LSND region below the KARMEN exclusion region and to the
left of the Bugey exclusion region gives the candidate oscillation
parameters.  This figure was provided by Bill Louis.}
\end{figure}

\subsection{Atmospheric Neutrinos}
When high energy
cosmic rays strike the earth's
atmosphere a multitude of secondary particles are produced, most of
which travel at nearly the speed of light in the same direction as the
incident cosmic ray.  Many of the secondaries are pions
and kaons, which decay as described above into electrons, muons,
and electron and muon neutrinos and antineutrinos.  These neutrinos
reach and pass through the earth.  The fluxes are large: about a
hundred such cosmic ray-induced neutrinos pass through each of us every
second.  Yet because these particles react weakly, only one
interaction is expected per human body every thousand years!  Thus a
considerably larger target is required for a reasonable event rate.
In SuperKamiokande, the massive 50,000 ton water detector that
replaced the original 3000 ton Kamiokande detector, one event
occurs every 90 minutes.  The energies of these neutrinos
(typically 1 GeV) are sufficiently high to produce either electrons
or muons, depending on the neutrino flavor.
As these charged particles pass through the water, they
produce Cerenkov radiation.  However, the Cerenkov ring produced
by an energetic electron is more diffuse than the relatively clean
ring of a muon.  This allows the experimenters to distinguish
electrons from muons with about 98\% accuracy.  Since the charged
lepton tends to travel in the same direction as the incident
neutrino, the experimenters can thus deduce both the flavor
and the direction of neutrinos that react in the water.

A decade ago it was already apparent that atmospheric neutrino
rates seen in existing detectors were anomalous.
Using known cross sections and decay rates theorists had predicted
about twice as many muon
neutrinos as electron neutrinos from cosmic ray events.
For example, we mentioned previously that a $\pi^+$ decays into
an $e^+$, a $\nu_e$, a $\nu_\mu$, and a $\bar{\nu}_\mu$.
That is, two muon neutrinos are produced, but only one $\nu_e$.
However, most of the early atmospheric neutrino experiments
found the electron-to-muon ratio from neutrino reactions to be approximately
unity.  The very precise measurements made with SuperKamiokande appear
to show that the ratio has this unexpected value because of a deficit
in muon-like events---the electron event rate is about as expected.
The muon deficit has a strong zenith angle dependence, with the
largest suppression associated with atmospheric neutrinos
coming from below, {\it e.g.}, originating on the opposite side of
the earth.  Such a dependence of the muon-to-electron ratio
on distance is a signature of neutrino oscillations, as we have
noted.  The most plausible interpretation of the SuperKamiokande
data \cite{atn} is that atmospheric $\nu_\mu$'s are oscillating into
$\nu_\tau$'s, which are not observed because the $\nu_\tau$'s are
too low in energy to produce $\tau$'s in SuperKamiokande.  The strong
suppression
in the $\nu_\mu$ flux is characteristic of maximal mixing
($\theta \sim \pi/4$), while the zenith angle dependence
indicates that the oscillation length is comparable to the
earth's diameter.  The corresponding $\delta m^2$ is $\sim 2 \cdot
10^{-3}$ eV$^2$.  As many theoretical models predict the
$\nu_\tau$ to be significantly heavier than the $\nu_\mu$,
this suggests a $\nu_\tau$ mass of $\sim 0.05$ eV.
The quality of the SuperKamiokande data -- the statistical error
on the muon-to-electron event rate is well below 10\% and there
is remarkable consistency between the sub-GeV and multi-GeV
data sets and between the fully- and partially-contained data sets --
provides a powerful argument that oscillations have been
observed.  Because the zenith-angle dependence shows that the
$\nu_\mu$ flux depends on distance, the atmospheric data provide
direct proof of oscillations.  Thus this may be our strongest
evidence for massive neutrinos and for the incompleteness of
the standard model.

There is another remarkable aspect of the atmospheric neutrino results.
If neutrinos are massive, there must be some reason that their
masses are so much lighter than those of all the more familiar
quarks and leptons.  In fact, a lovely explanation is provided
in many proposed extensions of the standard model that again returns to the
idea
that neutrinos are special because they can have both Dirac and Majorana
masses.  The
explanation is called the seesaw mechanism \cite{seesaw}.  It
predicts that the neutrino mass is
\begin{equation}
m_\nu = m_D ({m_D \over m_R})
\end{equation}
where $m_D$ is a Dirac mass often equated to the mass of the
corresponding quark or charged lepton, while
$m_R$ is a very heavy Majorana mass
associated with interactions at energies far above the reach
of existing accelerators.  This seesaw mechanism arises naturally
in models with both Dirac and Majorana masses.
In the case of the $\nu_\tau$ we
concluded from the atmospheric neutrino data that its mass
might be $\sim$ 0.05 eV.  A reasonable choice for $m_D$ is
the mass of the corresponding third generation quark, the
top quark, $m_D \sim$ 200 GeV.  It follows
that $m_R \sim 10^{14}$ GeV!  Thus tiny neutrino masses might
be our window on physics at enormous energy scales.  This large
mass $m_R$ is interesting because there is an independent argument,
based on observations that the weak, electromagnetic, and
strong interactions would all have approximately the same
strength at $\sim 10^{16}$ GeV, that suggests a very similar
value for the ``grand unification scale."
This has led many in the community to hope that
the pattern of neutrino masses now being discovered may
help us probe the structure of the theory that lies beyond
the standard model \cite{wilczek}.

\subsection{Solar Neutrinos}
The thermonuclear reactions occurring in its
core make the sun a marvelous source of neutrinos of a single
flavor, $\nu_e$.  The standard solar model -- really, the standard
model of main sequence stellar evolution -- allows us to predict
the flux and spectrum of these neutrinos.  The standard solar
model makes four basic assumptions: \\
$\bullet$ The sun evolves in hydrostatic equilibrium, maintaining a local
balance between the gravitational force and the pressure
gradient.  To describe this condition, one must specify the
equation of state as a function of temperature, density, and
composition. \\
$\bullet$ Energy is transported by radiation and convection.  While the
solar envelope is convective, radiative transport dominates in
the core region where thermonuclear reactions take place.
The opacity depends sensitively on the solar composition,
particularly the abundances of heavier elements. \\
$\bullet$ Solar energy is produced by thermonuclear reaction chains
in which four protons are converted to $^4$He
\begin{equation}
4 \mathrm{p} \rightarrow {}^4\mathrm{He} + 2e^+ + 2\nu_e
\end{equation}
The standard solar model predicts that 98\% of these reactions
occur through the pp chain illustrated in Figure 8, with
the CNO cycle accounting for the remainder.  The sun is a
slow reactor, characterized by a relatively low core temperature
$T_c \sim 1.5 \cdot 10^7$ K.  Thus Coulomb barriers tend to
suppress the rates of reactions involving higher Z nuclei,
an effect we will see reflected in the neutrino fluxes given below. \\
$\bullet$ The model is constrained to produce today's solar radius,
mass, and luminosity.  An important assumption of the standard
solar model is that the sun was highly convective, and therefore
uniform in composition, when it first entered the main
sequence.  It is furthermore assumed that the surface
abundances of metals (nuclei heavier than He) were undisturbed by
the sun's subsequent evolution, and
thus provide a record of the initial core metallicity.
The remaining parameter is the initial $^4$He/H ratio,
which is adjusted until the model reproduces the known solar
luminosity at the sun's present age, 4.6 billion years. \\

\begin{figure}
\psfig{bbllx=0.5cm,bblly=4.0cm,bburx=18cm,bbury=18.5cm,figure=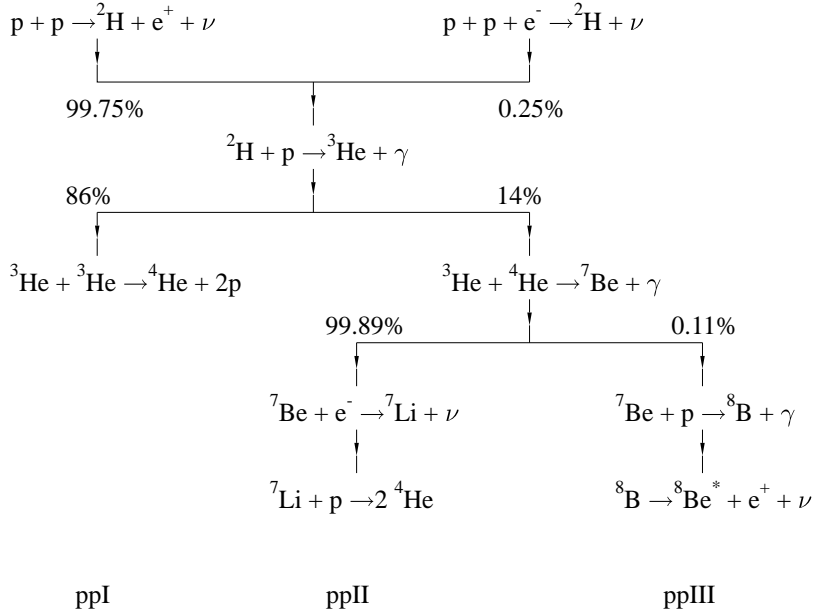,height=3.3in}
\caption{The three cycles comprising the pp chain.}
\end{figure}

Figure 8 shows that the pp chain is comprised of three
distinct cycles, each of which is tagged by a distinctive
neutrino.  The total rate of ppI+ppII+ppIII burning is
governed by the rate at which protons are consumed
\begin{equation}
\mathrm{p}+\mathrm{p} \rightarrow {}^2\mathrm{H} + e^+ + \nu_e
\end{equation}
a reaction which produces an allowed $\beta$ decay spectrum (that is, a
spectrum like that of Eq. \ref{eq:ttt})
of low-energy (0.42 MeV endpoint) $\nu_e$'s.  The ppII rate
is tagged by the distinctive neutrino lines from electron
capture on $^7$Be (0.86 and 0.36 MeV).  Finally the ppIII
cycle is tagged by the high-energy neutrinos
from the $\beta$ decay of $^8$B ($\sim$ 15 MeV endpoint).
The competition between these three cycles depends sensitively
on the solar core temperature $T_c$.  Thus the original
motivation for measuring solar neutrinos was to determine
the relative rates of the ppI, ppII, and ppIII cycles, from which the
core temperature could be deduced to an accuracy of a few
percent, thereby checking the standard solar model.

The neutrino flux predictions of the standard solar model are
summarized below \cite{bp98}.
\begin{equation}
\begin{array}{lll}
{\rm Reaction}&E_\nu^{\rm max}&{\rm Flux}(10^{10}/{\rm cm}^2/{\rm sec})\\
{\rm p+p}\rightarrow ^2{\rm H}+e^++\nu_e & 0.42 & 5.94\\
{}^7{\rm Be}+e^-\rightarrow{}^7{\rm Li}+\nu_e & 0.86(90\%)& 4.80\times
10^{-1}\\
  & 0.36(10\%)& \\
{}^8{\rm B}\rightarrow{}^8{\rm Be}^*+e^++\nu_e&14.06&5.15\times 10^{-4}
\end{array}
\end{equation}

The first experiment to test these predictions began more than
three decades ago with a detector placed a mile underground in the Homestake
Gold Mine in
Lead, South Dakota.  This great depth protects the detector
from all forms of cosmic radiation other than neutrinos.
Ray Davis, Jr., and his collaborators \cite{dav} filled the
detector with 615 tons
of the cleaning fluid perchloroethylene (C$_2$Cl$_4$)
in order to make use of the reaction
\begin{equation}
^{37}\mathrm{Cl}(\nu_e,e^-)^{37}\mathrm{Ar}
\end{equation}
The few atoms of the noble gas $^{37}$Ar produced in the tank
after a typical ($\sim$ two month) exposure could be recovered
quantitatively by a helium purge,
then counted via the subsequent electron capture reaction
$e^- + ^{37}$Ar $\rightarrow ^{37}$Cl + $\nu_e$, which has a
35-day half life.  The miniscule $^{37}$Ar production rate, less than
an atom every two days, has been measured to an accuracy of
better than 10\% by patient effort.
The deduced solar neutrino capture rate, 2.56 $\pm$ 0.16 $\pm$ 0.16
SNU (1 SNU = 10$^{-36}$ captures/atom/sec) is about 1/3 of the
standard solar model prediction.  As this reaction is primarily
sensitive to $^8$B (78\%) and $^7$Be (15\%) neutrinos, one
concludes that the sun is producing fewer high energy neutrinos
than expected.

Similar radiochemical experiments were done by the SAGE and
GALLEX collaborations \cite{gal} using a different target,
one containing $^{71}$Ga.  The special properties of this
nucleus include an unusually low threshold for the reaction
$^{71}$Ga($\nu_e,e^-)^{71}$Ge, leading to a large cross section for the capture
of low-energy pp neutrinos.
The resulting experimental capture rates are
66 $\pm$ 13 $\pm$ 6 and 76 $\pm$ 8 SNU for the SAGE and
GALLEX detectors, respectively, which can be compared to the
standard solar model prediction of $\sim$ 130 SNU.  Most
important, as the pp flux is directly constrained by the
rate of hydrogen burning and thus by the observed solar
luminosity in all steady-state solar models, there is a
minimum theoretical value for the capture rate of 79 SNU, given
standard model weak interaction physics.  With this assumption,
it appears that there is virtually no contribution from ppII
and ppIII cycle neutrinos.

The remaining experiments, Kamiokande II/III and the ongoing
SuperKamiokande \cite{kam}, exploit water Cerenkov detectors to view
solar neutrinos on an event-by-event basis.  Just as described
in our atmospheric neutrino discussion, the scattering of
high energy $^8$B neutrinos produces recoil electrons and thus
Cerenkov radiation that can be recorded in the surrounding
phototubes.  The correlation of the electron direction with
the position of the sun is crucial in separating solar neutrino
events from background.  After 504 days of operations the rate
measured by SuperKamiokande, which under
current operating conditions is sensitive to neutrinos with
energies above $\sim$ 6 MeV, is consistent with an $^8$B
neutrino flux of (2.44$\pm0.05^{+0.09}_{-0.07}) \cdot 10^6$/cm$^2$sec.
This is about half of the standard solar model prediction.

If one combines the various experimental results and assumes that
the neutrino spectra are not being distorted by oscillations or
other new physics, the following pattern of fluxes emerges
\begin{eqnarray}
\phi (pp) & \sim & 0.9 \, \phi^{\rm {SSM}} (pp)\nonumber \\
\phi (^7{\rm {Be}}) & \sim & 0 \nonumber\\
\phi (^8 {\rm B}) & \sim & 0.4 \, \phi^{\rm {SSM}} (^8{\rm B})
\end{eqnarray}
where SSM stands for the standard solar model.
In fact, the preferred value of $\phi(^7$Be) turns out to be
negative (at 2-3 $\sigma$) in unconstrained fits.
A reduced $^8$B neutrino flux can be produced by lowering
the central temperature of the sun somewhat, as $\phi(^8$B)$\sim T_c^{18}$.
 However, such
an adjustment, either by varying the parameters of the standard solar model or
by
adopting some nonstandard physics, tends to push the $\phi (^7$Be)/$\phi(^8$B)
ratio to higher values rather than the low one required by the results above,
\begin{equation}
{\phi (^7{\rm{Be}}) \over \phi(^8 {\rm B})} \sim T_c^{-10}.
\end{equation}
Thus the observations seem difficult to reconcile with plausible
solar model variations: one observable, $\phi(^8$B), requires a cooler
core while a second, the ratio $\phi(^7$Be)/$\phi(^8$B), requires a hotter one.

These arguments seem to favor a more radical solution, one
involving new properties of neutrinos.  Originally the most
plausible such solution was neutrino oscillations of the type
we discussed in connection with atmospheric neutrinos and LSND.
For simplicity we consider the mixing of the $\nu_e$ with a
single second flavor, the $\nu_\mu$.  This will clearly alter
the expectation for solar neutrino experiments, as the
Homestake and SAGE/GALLEX experiments cannot detect $\nu_\mu$'s,
while SuperKamiokande detects $\nu_\mu$'s with reduced
efficiency.  (The cross section for $\nu_\mu$ scattering off
electrons is about 1/6 that for $\nu_e$ scattering.)
The $\nu_e$ survival probability at the earth is
\begin{equation}
p_{\nu_e}(x) = 1 - \sin^2(2\theta) \sin^2({\delta m^2 c^4 x
\over 4 \hbar c E}) \sim 1 - {1 \over 2} \sin^2(2\theta)
\end{equation}
where $\theta$ is the $\nu_e-\nu_\mu$ mixing angle.  (This is
just our earlier result of Eq. \ref{eq:osc}, with the replacement
$p \sim E$ and with the factors of $c$ and $\hbar$ reinserted.)
The result on the right is appropriate if the oscillation
length $L_0 = 4\pi \hbar c E/\delta m^2 c^4$ is much smaller
that the earth-sun distance $x$.  In that case (for a broad
spectrum such as the $^8$B neutrinos) the oscillatory factor
averages to 1/2.  For $^8$B neutrinos this averaging is
appropriate if $\delta m^2 c^4$ exceeds $10^{-9}$eV$^2$.

Even if the mixing angle were maximal --- $\theta \sim \pi/4$ ---
such vacuum oscillations would then produce a factor of two
suppression in the neutrino flux, not the factor of three
indicated by the $^{37}$Cl result.  Historically large mixing
angles were viewed as contrived, as the similar Cabibbo angle
describing the weak mixing of quarks is quite small.  The
recent atmospheric neutrino results, which strongly favor
large mixing angles, have weakened this argument.  Nevertheless,
one of the marvelous properties of our sun is that it can
greatly enhance oscillations even if mixing angles are quite
small.  We now turn to describing this effect, which is
known as the Mihkeyev-Smirnov-Wolfenstein mechanism.

The starting point is a slight generalization of our vacuum
neutrino oscillation discussion.  Previously we discussed the
case where our initial neutrino had a definite flavor.  But
we could have considered the somewhat more general case
\begin{equation}
 |\nu(t=0)\rangle = a_e(t=0) |\nu_e \rangle + a_\mu(t=0)
|\nu_\mu \rangle .
\end{equation}
Exactly as before we could expand this wave function in terms of
the mass eigenstates, which propagate simply, to find (this
takes a bit of algebra)
\begin{equation}
i {d \over dx} \left( \matrix { a_{\textstyle e} \cr
a_{\textstyle \mu} \cr} \right) = {1 \over 4E} \left ( \matrix{
- \delta m^2 \cos 2 \theta_{\textstyle v}
~~~~~~~~~~~\delta m^2\sin
2\theta_{\textstyle v} \cr
\delta m^2\sin 2 \theta_{\textstyle v} ~~~~~~~~~~~
\delta m^2
\cos 2\theta_{\textstyle v} \cr} \right) \left( \matrix {
a_{\textstyle e} \cr
a_{\textstyle \mu} \cr} \right) .
\end{equation}
Note that the common phase has been ignored: it can be absorbed
into the overall phase of the coeficients $a_e$ and $a_\mu$,
and thus has no consequence.  We have also labeled the mixing angle as
$\theta_{\textstyle v}$, to emphasize that it is the vacuum value,
and equated $x = t,$ that is, set $c$ = 1.

The view of neutrino oscillations changed
when Mikheyev and Smirnov~\cite{ms} showed in 1985 that the
density dependence of the neutrino effective mass, a phenomenon
first discussed by Wolfenstein~\cite{wolf} in 1978, could greatly enhance
oscillation probabilities: a $\nu_e$ is adiabatically transformed
into a $\nu_\mu$ as it traverses a critical density within the sun.
It became clear that the sun was not only an excellent
neutrino source, but also a natural regenerator for cleverly
enhancing the effects of flavor mixing.

The effects of matter alter our neutrino evolution equation in
an apparently simple way
\begin{equation}
i {d \over dx} \left( \matrix { a_{\textstyle e} \cr
a_{\textstyle \mu} \cr} \right) = {1 \over 4E} \left ( \matrix{
2E \sqrt2 G_F \rho(x) - \delta m^2 \cos 2 \theta_{\textstyle v}
~~~~~~\delta m^2\sin
2\theta_{\textstyle v} \cr
\delta m^2\sin 2 \theta_{\textstyle v} ~~~ -2E \sqrt2 G_F \rho(x) +
\delta m^2
\cos 2\theta_{\textstyle v} \cr} \right) \left( \matrix {
a_{\textstyle e} \cr
a_{\textstyle \mu} \cr} \right) \label{eq:mm}
\end{equation}
where $\rho(x)$ is the solar electron density.
The new contribution to the diagonal elements, $2 E \sqrt2 G_F \rho(x)$,
represents the effective contribution to $m^2_\nu$  that arises
from neutrino-electron scattering.  The indices of refraction
of electron and muon neutrinos differ because the former
scatter by charged and neutral currents, while the latter
have only neutral current interactions: the sun contains electrons
but no muons.  The difference in
the forward scattering amplitudes determines the density-dependent
splitting of the diagonal elements of the new matter equation.

It is helpful to rewrite this equation in a basis consisting of the light and
heavy
local mass eigenstates (i.e., the states that diagonalize the right-hand side
of the equation),
\begin{eqnarray}
|\nu_L (x)\rangle &=& \cos \theta (x)|\nu_e\rangle - \sin \theta
(x)|\nu_\mu\rangle \nonumber \\
|\nu_H(x)\rangle &=& \sin \theta (x)|\nu_e\rangle + \cos \theta (x)|\nu_\mu
\rangle .
\end{eqnarray}
The local mixing angle is defined by
\begin{eqnarray}
\sin 2 \theta (x)  &=& {\sin 2 \theta_{\textstyle v} \over \sqrt{X^2 (x) +
\sin^2
2\theta_{\textstyle v}}} \nonumber \\
\cos 2\theta (x)  &=& {-X (x) \over \sqrt{X^2 (x) + \sin^2 2\theta_{\textstyle
v}}}
\end{eqnarray}
where $X(x) = 2 \sqrt2G_F \rho(x) E/\delta m^2 - \cos 2\theta_{\textstyle v}$.
Thus
$\theta(x)$ ranges from $\theta_{\textstyle v}$ to $\pi/2$ as the density
$\rho(x)$ goes
from 0 to $\infty$.

If we define
\begin{equation}
|\nu (x) \rangle = a_H(x)|\nu_H(x)\rangle + a_L(x)|\nu_L(x)\rangle,
\end{equation}
the neutrino propagation can be rewritten in terms of the local
mass eigenstates
\begin{equation}
i {d \over dx} \pmatrix{
a_H \cr
a_L \cr} = \pmatrix {
\lambda(x) & i \alpha (x) \cr
-i \alpha (x) & - \lambda (x) \cr }
\pmatrix
{a_H \cr
a_L }
\end{equation}
with the splitting of the local mass eigenstates determined by
\begin{equation}
2 \lambda (x) = {\delta m^2 \over 2E} \sqrt{X^2 (x) + \sin^2 2
\theta_{\textstyle v}}
\end{equation}
and with mixing of these eigenstates governed by the density gradient
\begin{equation}
\alpha (x) = \left({E \over \delta m^2}\right)
 \, {\sqrt2 \, G_F {d \over dx}
\rho(x)
\sin 2 \theta_{\textstyle v} \over X^2 (x) + \sin^2 2 \theta_{\textstyle v}}.
\end{equation}
The results above are quite interesting: the local mass eigenstates
diagonalize the matrix if the density is constant, that is, if $\alpha$ = 0.
 In such a limit,
the problem is no more complicated than our original vacuum
oscillation case, although our mixing angle is changed because of
the matter effects.  But if the density is not constant, the
mass eigenstates in fact evolve as the density changes.  This
is the crux of the MSW effect.
Note that the splitting achieves
its minimum value, ${\delta m^2 \over 2E} \sin 2 \theta_v$, at a critical
density $\rho_c =
\rho (x_c)$
\begin{equation}
2 \sqrt2 E G_F \rho_c = \delta m^2 \cos 2 \theta_v
\end{equation}
that defines the point where the diagonal elements of the matrix in Eq.
\ref{eq:mm} cross.

Our local-mass-eigenstate form of the propagation equation can be trivially
integrated if the splitting of the diagonal
elements is
large compared to the off-diagonal elements,
so that the effects of $\alpha(x)$ can be ignored
\begin{equation}
\gamma (x) = \left|{\lambda (x) \over \alpha (x)}\right| = {\sin^2
2\theta_{\textstyle v} \over \cos
2\theta_{\textstyle v}} \, {\delta m^2 \over 2 E} \, {1 \over |{1 \over \rho_c}
{d \rho (x) \over
dx}|} {[X (x)^2 + \sin^2 2\theta_v]^{3/2} \over \sin^3 2\theta_v} \gg 1,
\end{equation}
a condition that becomes particularly stringent near the crossing point,
where $X(x)$ vanishes,
\begin{equation}
\gamma_c = \gamma (x_c) = {\sin^2 2\theta_v \over \cos 2\theta_v} \, {\delta
m^2 \over 2 E} \, {1 \over \left|{1 \over \rho_c} {d \rho (x) \over dx}|_{x =
x_c}\right|}
\gg 1.
\end{equation}
The resulting adiabatic electron neutrino survival probability~\cite{bethe},
valid when
$\gamma_c \gg 1$, is
\begin{equation}
p^{\rm adiab}_{\nu_e} = {1 \over 2} + {1 \over 2} \cos 2 \theta_v \cos 2
\theta_i
\end{equation}
where $\theta_i = \theta (x_i)$ is the local mixing angle at the density where
the neutrino was produced.  (So if $\theta_v \sim 0$ and if the
starting solar core density is sufficiently high so that
$\theta_i \sim \pi/2$, $p^{\rm adiab}_{\nu_e} \sim 0$.)

\begin{figure}[htb]
\psfig{bbllx=1.2cm,bblly=2.0cm,bburx=18cm,bbury=14.5cm,figure=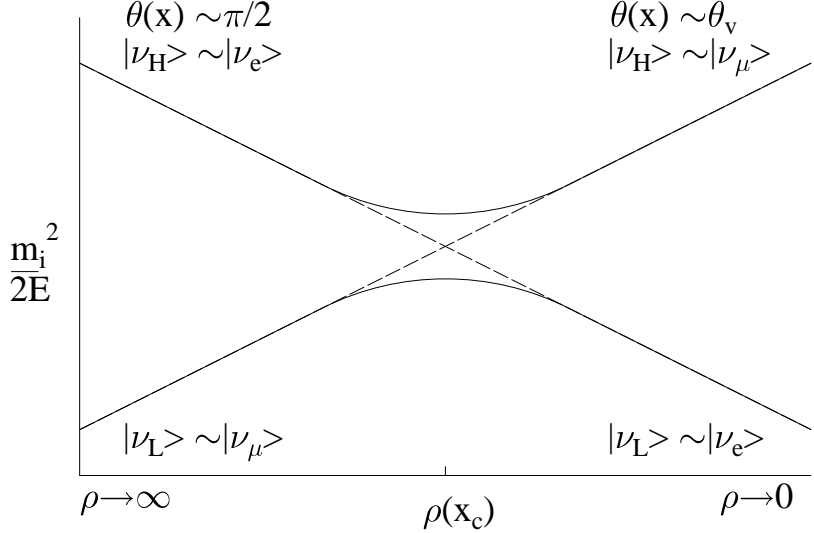,height=3.0in}
\caption{Schematic illustration of the MSW crossing.  The dashed
lines correspond to the electron-electron and muon-muon diagonal
elements of the $m_\nu^2$ matrix in the flavor basis.  Their
intersection defines the level-crossing density $\rho_c$.
The solid lines are the trajectories of the light and heavy
local mass eigenstates.  If the electron neutrino is produced
at high density and propagates adiabatically, it will follow
the heavy-mass trajectory, emerging from the sun as a $\nu_\mu$.}
\end{figure}

The physical picture behind this derivation is illustrated
in Figure 9.  One makes the usual assumption that, in vacuum,
the $\nu_e$ is almost identical to the light mass eigenstate,
$\nu_L(0)$, i.e., $m_1 < m_2$ and $\cos \theta_v \sim$ 1.  But as the density
increases,
the matter effects make the $\nu_e$ heavier than the $\nu_\mu$, with $\nu_e
\to \nu_H (x)$  as $\rho(x)$ becomes large.  The special property of
the sun is that it produces $\nu_e$'s at high density that then propagate to
the vacuum where they
are measured.  The adiabatic approximation tells us that if
initially $\nu_e \sim \nu_H (x)$, the neutrino will remain on the heavy
mass trajectory provided the density changes slowly.
That is, if the solar density gradient is sufficiently gentle,
the neutrino will emerge from the sun as the heavy vacuum
eigenstate, $ \sim \nu_\mu$.  This guarantees nearly complete conversion
of $\nu_e$'s into $\nu_\mu$'s, producing a flux that cannot be detected
by the Homestake or SAGE/GALLEX detectors.

Although it goes beyond the scope of this discussion, the case
where the crossing is nonadiabatic can also be handled in an
elegant fashion by following a procedure introduced by Landau and Zener
for similar atomic physics level-crossing problems.  The result \cite{haxpar}
is an oscillation probability valid for all $\delta m^2/E$ and
$\theta_v$
\begin{equation}
p_{\nu_e} = {1 \over 2} + {1 \over 2} \cos 2 \theta_v \cos 2 \theta_i ( 1 -
2 e^{-\pi \gamma_c/2})
\end{equation}
As it must by our construction, $p_{\nu_e}$ reduces to $p^{\rm
{adiab}}_{\nu_e}$ for $\gamma_c \gg$ 1.
When the crossing becomes nonadiabatic ({\it e.g.}, $\gamma_c \ll 1$ ),
the neutrino ``hops" to the light mass trajectory as it
reaches the crossing point, allowing the neutrino to
exit the sun as a $\nu_e$, i.e., no conversion
occurs.

Thus there are two conditions for strong
conversion of solar neutrinos:  there must be a level
crossing (that is, the solar core density must be sufficient
to render $\nu_e \sim \nu_H (x_i)$  when it is first
produced) and the crossing must be adiabatic.  The first
condition requires that $\delta m^2/E$ not be too large, and the
second $\gamma_c \gsim$ 1.  The combination of these two constraints,
illustrated in Figure 10, defines a triangle of interesting
parameters in the ${\delta m^2 \over E} - \sin^2 2\theta_v$ plane, as Mikheyev
and Smirnov
found by numerical
integration.  A remarkable feature of this triangle
is that strong $\nu_e \to \nu_\mu$ conversion can occur for very small
mixing angles $(\sin^2 2 \theta \sim10^{-3}$), unlike the vacuum case.

\begin{figure}[htb]
\psfig{bbllx=-2.5cm,bblly=0.0cm,bburx=15cm,bbury=22.0cm,figure=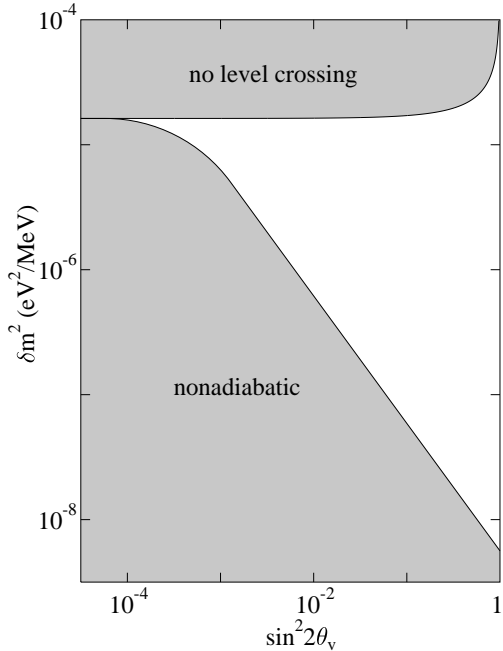,height=3.6in}
\caption{MSW conversion for a neutrino produced at the sun's
center.  The upper shaded region indices thoses $\delta m^2/E$
where the vacuum mass splitting is too great to be overcome
by the solar density.  Thus no level crossing occurs.  The
lower shaded region defines the region where the level crossing
is nonadiabatic ($\gamma_c$ less than unity).  The unshaded
region corresponds to adiabatic level crossings where strong
$\nu_e \rightarrow \nu_\mu$ will occur.}
\end{figure}

One can envision superimposing on Figure 10 the spectrum of solar neutrinos,
plotted as a
function of ${\delta m^2 \over E}$ for some choice of $\delta m^2$.
Since Davis sees {\it some} solar neutrinos, the solutions must
correspond to the boundaries of the triangle in Figure 10.  The horizontal
boundary indicates the maximum ${\delta m^2 \over E}$ for which the sun's
central density is sufficient to cause a level crossing.  If a spectrum
properly straddles this boundary, we obtain a result consistent with the
Homestake experiment in which low energy neutrinos (large 1/E) lie above the
level-crossing boundary (and thus remain $\nu_e$'s), but the high-energy
neutrinos (small 1/E) fall within the unshaded region where strong conversion
takes place.  Thus such a solution would mimic nonstandard solar models in
that only the $^8$B neutrino flux would be strongly suppressed.  The diagonal
boundary separates the adiabatic and nonadiabatic regions.  If the spectrum
straddles this boundary, we obtain a second solution in which low energy
neutrinos lie within the conversion region, but the high-energy neutrinos
(small 1/E) lie below the conversion region and are characterized by $\gamma
\ll 1$ at the crossing density.  (Of course, the boundary is not a sharp one,
but is characterized by the Landau-Zener exponential).  Such a nonadiabatic
solution is quite distinctive as the flux of  pp neutrinos, which is
strongly constrained in the standard solar model and in any steady-state
nonstandard model by the solar luminosity, would now be sharply reduced.
Finally, one can imagine ``hybrid" solutions where the spectrum straddles both
the level-crossing (horizontal)
boundary and the adiabaticity (diagonal) boundary for small $\theta$,
thereby reducing the $^7$Be neutrino flux more than either the
pp or $^8$B fluxes.

Remarkably, this last possibility seems quite consistent with the
experiments we have discussed.  In fact, a nearly perfect fit to the data
results from choosing $\delta m^2 \sim 5 \cdot 10^{-6}$ eV$^2$
and $\sin^22\theta_v \sim 0.006$.  As the $\delta m^2$ is quite
different from that found in the atmospheric neutrino results,
the mixing seen in solar neutrinos is distinct from that seen in
atmospheric neutrinos, consistent with our attribution of the
former to $\nu_e \rightarrow \nu_\mu$ oscillations
and the latter to $\nu_\mu \rightarrow \nu_\tau$.

The argument that the solar neutrino problem must be due to
neutrino oscillations is quite strong, but not as compelling as in the
case of atmospheric neutrinos.  Our conclusions
follow from combining the results of several experiments,
and not from direct observation of new physics, such as the
zenith angle dependence in the atmospheric results.
For this reason there is great interest in a new experiment
now being readied in the Creighton nickel mine in Sudbury, Ontario,
6800 feet below the surface.  The Sudbury Neutrino Observatory
(SNO) \cite{sno} has a central acrylic vessel filled with one kiloton
of very pure (99.92\%) heavy water, surrounded by a shield of
7.5 kilotons of ordinary water.  SNO can detect neutrinos through
the charged current reaction
\begin{equation}
\nu_e + d \rightarrow p + p + e^-
\end{equation}
as well as through the neutral current reaction
\begin{equation}
\nu_x(\bar{\nu}_x) + d \rightarrow \nu_x(\bar{\nu}_x) + p + n
\end{equation}
Thus SNO offers the exciting possibility of comparing the solar
flux in $\nu_e$'s with that in all flavors, thereby providing
a definitive test of flavor oscillations.  The electrons from
the first reaction above will be detected by the Cerenkov
light they generate: SNO's central vessel is surrounded by
9800 phototubes.  The neutrons produced in the neutral current
reaction can be detected using either the $(n,\gamma)$
reaction on salt dissolved in the heavy water or proportional counters
exploiting the $^3$He(n,p)$^3$H reaction.  SNO should begin taking data in
summer, 1999.

\section{Conclusions}

In this article we have summarized some of the basic ideas of neutrino physics.
Neutrinos come in three
species or flavors---$\nu_e,\nu_\mu,\nu_\tau$---with their corresponding
antiparticles.  We have discussed the unresolved problem of
the nature of these antiparticles: are $\nu$'s and $\bar{\nu}$'s distinquished
by some additive quantum number (Dirac neutrinos), or do they
instead correspond to the two projections of
opposite handedness of the same state (Majorana neutrinos)?
We reviewed direct mass measurements, which have not yet
yielded evidence for nonzero masses, and the reasons that the
minimal standard model cannot accommodate massive neutrinos.
Yet we have seen strong though indirect evidence for massive
neutrinos in three classes of neutrino oscillation experiments.
In an accelerator experiment---LSND---an unexpected flux of electron
antineutrinos has been attributed to $\bar{\nu}_\mu \rightarrow
\bar{\nu}_e$ oscillations.
Over the past 15 years a series of experiments have been carried
out on the neutrinos produced when high energy cosmic rays
interact in the upper atmosphere.  The increasing evidence of
an anomaly has culminated with the SuperKamiokande measurements
that confirm a deficit in the muon neutrino flux, and find
a zenith angle dependence directly indicating oscillations.
When constraints from reactor and accelerator experiments are
taken into account, the explanation for the atmospheric neutrino
anomaly appears to be $\nu_\mu \rightarrow \nu_\tau$ oscillations
corresponding to a nearly maximal mixing angle.  Finally,
we discussed the evidence for a deficit in the flux of solar
electron neutrinos and the difficulty in attributing this deficit
to uncertainties in the solar model.  Again the hypothesis of
neutrino oscillations accounts for the observations, with
one attractive possibility being a $\nu_e \rightarrow \nu_\mu$
oscillation enhanced by matter effects within the sun.

\begin{figure}
\psfig{bbllx=3.0cm,bblly=7.0cm,bburx=15cm,bbury=17cm,figure=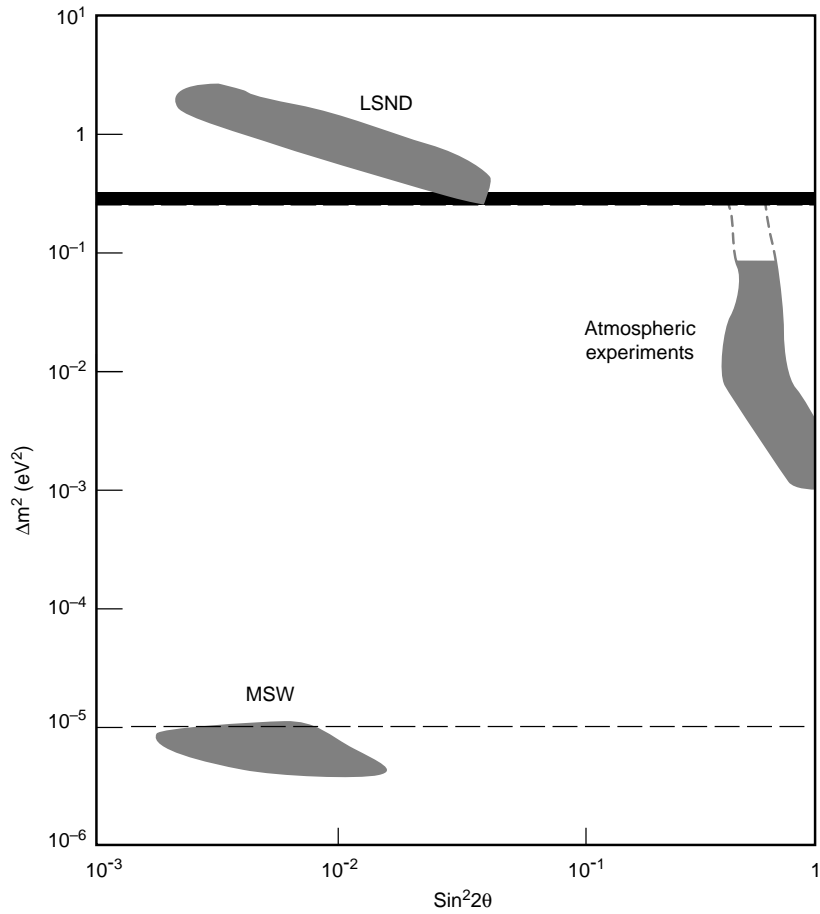,height=3.7in}
\caption{One set of $\delta m^2$ and $\sin^2 2\theta$ regions 
that could account for the solar neutrino, atmospheric neutrino,
and LSND results.  This figure was provided by Vern Sandberg.}
\end{figure}

One popular choice for the mass differences and mixing angles
needed to account for these observations is shown in Figure 11.
The pattern that emerges is not, unfortunately, compatible
with the simplest scenario of three mixed neutrinos: three
distinct $\delta m^2$ values are required, while a
three-neutrino scenario provides only two.  This could mean
that one (or more) of the experiments we have discussed has
been misinterpretted.  Alternatively, it could indicate the
existence of additional light neutrinos that have so far avoided
detection because they are sterile, lacking the usual standard
model interactions.  In any case, it is clear that the possibilities
summarized in Figure 11 will require much more study in a new generation
of heroic neutrino experiments.  We have noted two efforts that
made provide important results in the not too distant future:
SNO will soon tell us whether there exists
a $\nu_\mu$ or $\nu_\tau$ component in the solar neutrino flux,
while a new Fermilab experiment will test the LSND conclusions.
Thus perhaps the picture will become far clearer in the next few years.
In the meantime, neutrinos will remain in the news.  We have tried
to illustrate how the many profound issues connected with
neutrinos can be readily appreciated with only simple quantum
mechanics.  It is our hope, therefore, that the developing
discoveries in neutrino physics---which may show the way to
the next standard model of particle physics---will be accessible
to and followed by beginning students of physics.
We hope the material presented here, which is suitable for
beginning courses in quantum mechanics and modern physics,
will help make this possible.

\begin{center}
{\bf Acknowledgement}
\end{center}
This work is supported in part by the National Science Foundation (BRH) and
by the Department of Energy (WCH).  In addition BRH would like to acknowledge
the
support of the Alexander von Humboldt Foundation and the hospitality of
Forschungszentrum J\"{u}lich.  We also thank Bill Louis and Vern Sandberg
for providing Figures 7 and 11.


\begin{thebibliography}{99}
\bibitem{rabi} Often quoted is Rabi's remark about the discovery of
the muon---``Who ordered that?''
\bibitem{ftnt} Of course, decay of the $\pi^+$ to an electron and
neutrino is also possible but is highly suppressed for helicity
reasons, as we shall discuss below.
\bibitem{rth} In priniciple there could be right-handed neutrinos
present, but they could not interact and hence we would not be
aware of their presence.
\bibitem{con} The constant in front of the neutrino expression differs
from the corresponding photon number because of the use of Fermi
rather than Bose statistics
\begin{equation}
\int_0^\infty dz{z^3\over e^z-1}={\pi^4\over
15}\rightarrow\int_0^\infty dz{z^3\over e^z+1}={7\pi^4\over 120}.
\end{equation}
\bibitem{anti} The $\bar{\nu}_e$'s come from both the fraction of $\mu^-$'s
that decay
via $\mu^-\rightarrow e^-+\bar{\nu}_e+\nu_\mu$ and
from the small percentage of negatively charged pions that decay via
$\pi^-\rightarrow
e^-+\bar{\nu}_e$.
\end{thebibliography}
\end{document}